\documentclass{JHEP3}

\usepackage{graphicx}
\usepackage{mathrsfs}
\usepackage{amsmath, amsthm, amssymb}

\newcommand{\be}{\begin{equation}}
\newcommand{\ee}{\end{equation}}
\newcommand{\beq}{\begin{equation}}
\newcommand{\eeq}{\end{equation}}
\newcommand{\bea}{\begin{eqnarray}}
\newcommand{\eea}{\end{eqnarray}}

\def\P{{\cal P}}
\def\muth{\mu^{th}}

\newcommand{\gsim}{\lower.7ex\hbox{$\;\stackrel{\textstyle>}{\sim}\;$}}
\newcommand{\lsim}{\lower.7ex\hbox{$\;\stackrel{\textstyle<}{\sim}\;$}}

\title{Histogram comparison as a powerful tool for the search of new physics at LHC. Application to CMSSM}

\author{Maria Eugenia Cabrera\\
        Instituto de F\'isica Te\'orica, IFT-UAM/CSIC \\
        U.A.M., Cantoblanco, \\
        28049 Madrid, Spain}

\author{J. Alberto Casas\\
        Instituto de F\'isica Te\'orica, IFT-UAM/CSIC \\
        U.A.M., Cantoblanco, \\
        28049 Madrid, Spain }

\author{Vasiliki A. Mitsou\\
        Instituto de F\'isica Corpuscular, IFIC-UV/CSIC \\
        Valencia, Spain }

\author{Roberto Ruiz de Austri\\
        Instituto de F\'isica Corpuscular, IFIC-UV/CSIC \\
        Valencia, Spain }

\author{Juan Terr\'on\\
Departamento de Fisica Te\'orica C-15, \\ 
Universidad Aut\'onoma de Madrid, \\
Madrid, Spain }

\abstract{\small

We propose a rigorous and effective way to compare experimental and theoretical histograms, incorporating the different sources of statistical and systematic uncertainties. This is a useful tool to extract as much information as possible from the comparison between experimental data with theoretical simulations, optimizing the chances of identifying New Physics at the LHC. We illustrate this by showing how a search in the CMSSM parameter space, using Bayesian techniques, can effectively find the correct values of the CMSSM parameters by comparing histograms of events with multijets + missing transverse momentum displayed in the effective-mass variable. The procedure is in fact very efficient to identify the true supersymmetric model, in the case supersymmetry is really there and accessible to the LHC.

}

\preprint{IFT-UAM/CSIC-11-67}

\begin{document}

\section{Introduction}
\label{sec:intro}

The LHC is already probing new physics beyond the reach of past experiments. At any stage of this enterprise, i.e. with the available data at any time, there are two main questions to address: 1) Is there any signal of New Physics (NP)? and 2) In the positive case, which NP is it? In order to optimize the answer to these questions there is an intense activity to explore assorted strategies for the search of NP. The task is challenging, due in part to the fact that LHC data, though very rich, are not as clean as those from an $e^+ e^-$ collider. Besides, the theoretical calculations are also subject to great uncertainties and rely to some extent on Monte Carlo simulations.

Most of the LHC data can be organized in form of histograms with number of
events of a certain kind (e.g. those presenting multijets + missing transverse
momentum) displayed in different variables \cite{Ball:2007zza, Aad:2009wy,
  alphat}: $M_{\rm eff}$, $p_T^{\rm miss}$, $\alpha_T$, etc. In many cases the comparison with the simulations is done just by comparing the total number of events after performing different cuts in the variables involved. In this way, both ATLAS and CMS have already posed meaningful bounds \cite{atlas, cms} on NP scenarios, in particular on the simplest supersymmetric model, the so-called Constrained Minimal Supersymmetric Standard Model (CMSSM) \cite{Kane:1993td}. More precisely, the most powerful bounds on the CMSSM have been obtained by considering events with several jets + missing transverse momentum. Somehow, the study of the total number events, choosing different cuts on the sets of data, amounts to partially compare the shapes of the experimental and the theoretical (simulated) histograms; although in a way which 
 is not optimal. 

As mentioned above, even if we are quite sure to have a signal of NP, we face the problem of identifying the model producing such signal. Of course the variety of scenarios of NP is enormous, which makes the job very complex. Even playing in the framework of a given scenario, such as the CMSSM, the sole study of the number of events of a certain kind is not enough to determine the parameters of the model, due to the existence of big degeneracies in such determination. Again, this situation can be improved by probing different cuts in the sets of data. But, once more, this is not an optimized way of comparing theory and experiment, since the richness of the data is not completely exploited. 

The goal of this paper is to propose an effective and rigorous way to compare experimental and theoretical histograms, incorporating the different sources of uncertainty involved in the task. In our opinion, in an experiment with the characteristics of the LHC this is a useful tool to extract as much information as possible from the comparison between experimental data with theoretical simulations. We illustrate this usefulness by showing how a search in the CMSSM parameter space, using Bayesian techniques, can effectively find the correct values of the CMSSM parameters by comparing histograms of events with multijets + missing transverse momentum displayed in the $M_{\rm eff}$ variable. This procedure could be very efficient to identify the true supersymmetric model, in the case supersymmetry is really there and accessible to the LHC.

In section 2 we establish the notation and the statistical basis for the rigorous comparison between the experimental and the theoretical histograms. Section 3 is devoted to the incorporation of extra sources of uncertainty, in particular systematic ones. At the end of this section we give our final formula for the complete likelihood of a theoretical model by histogram comparison. In section 4 we illustrate the proposed technique by showing how a search in the constrained-MSSM parameter space, using Bayesian techniques, can effectively find the correct values of the MSSM parameters by comparing histograms of events with multijets + missing transverse momentum displayed in the effective-mass variable. But, of course, the technique can be applied to any scenario of new physics. Our conclusions are summarized in section 5. Finally, in the Appendix we show how our final formula for histogram-comparison is (slightly) modified when the effective luminosity of the theoretical simulation is not the same as the experimental one. 

\section{Comparison of histograms. Statistical uncertainties}
\label{sec:histograms}

\subsection{Basic Ingredients and Notation}

Suppose we have experimental data, e.g. multijet + missing transverse momentum events at LHC, organized in an histogram upon some variable $M$, e.g. the effective mass of the events, as defined in ref. \cite{Aad:2009wy}. Let us call $K$ the number of bins of the histogram. Each bin corresponds to a (central) value of the effective mass, $M_i$. We will denote the bin contents (number of events for each $M_i$)
by $v_i$. The total number of events is $v=\sum_i v_i$.

Leaving apart for the moment all sources of systematic uncertainties, the probability that the experiment produces the actual data, $v_i$, is given by a Poisson distribution
\bea
\label{poiss_vi_nu}
\P(v_i)=\prod_{i=1}^{K}\ \frac{\nu_i^{v_i}}{v_i!}e^{-\nu_i} ,
\eea
where $\nu_i$ are the expected values (or ``means") of the distribution. The values of $\nu_i$ are in principle calculable (at some degree of precision) provided we knew the theory responsible for them, e.g. the Standard Model. But we are precisely trying to uncover unknown NP, therefore $\nu_i$ are unknown. 

On the other hand, working within a scenario of NP defined by some parameters,
$\theta_a$ (for example the parameters of the CMSSM), we can in principle calculate the means 
under, supposedly, the same conditions of energy and luminosity as the experiment. We will denote $\mu_i$ these theoretical means. Of course, $\mu_i$ depend on the point in the parameter space, i.e. the precise model under consideration. If the model is the true one, then $\nu_i=\mu_i$. This is the so-called ``null-hypothesis". The likelihood of a point of the parameter space is the corresponding probability of producing the observed data, $v_i$, under the null-hypothesis, i.e.
\bea
\label{poiss_vi}
\P(v_i)=\prod_{i=1}^{K}\ \frac{\mu_i^{v_i}}{v_i!}e^{-\mu_i} .
\eea
The likelihood is a crucial quantity to compare the viability of the different regions of the parameter space, both in frequentist and Bayesian analyses (see, e.g.~\cite{Trotta:2008qt}). In particular, in Bayesian analyses one is interested in determining the probability density of a point of the parameter space, $\theta_a$, given an experimental set of data (in our case, $v_i$). This is the so-called posterior probability density function (pdf), $p(\theta_a|{\rm data})$, which is given by the fundamental Bayesian relation 
\bea
\label{Bayes}
p(\theta_a|{\rm data})\ =\ p({\rm data}|\theta_a)\ p(\theta_a)\ \frac{1}{p({\rm data})}\ .
\eea
Here $p({\rm data}|\theta_a)$ is the above-mentioned likelihood, i.e. the probability of obtaining the observed data if the model defined by the $\theta_a$ parameters is the true one; while $p(\theta_a)$ is the prior, i.e. the ``theoretical" probability density that we assign a priori to the point in the parameter space; and $p({\rm data})$ is a normalization factor that ensures that the total probability is one. 

In order to compute the likelihood (\ref{poiss_vi}) we need the theoretical means, $\mu_i$. However, in practice one does not have at disposal a complete evaluation of $\mu_i$, but rather a simulation of the process using diverse computation codes. The results of the simulation can also be organized in an histogram with $K$ bins, associated with the same values of the effective mass, $M_i$. The bin contents of the simulation are denoted by $u_i$, with total number of events $u=\sum_i u_i$. Of course, the values of $u_i$ obey also a Poisson statistics
\bea
\label{poiss_ui}
\P(u_i)=\prod_{i=1}^{K}\ \frac{\mu_i^{u_i}}{u_i!}e^{-\mu_i} .
\eea
Here we have again left aside for the moment all sources of systematic uncertainties associated with the theoretical simulation.

\subsection{Computation of the likelihood}

As mentioned, usually the codes provide values for $u_i$, but not for
$\mu_i$. If we had enough computation time we could obtain a good evaluation
of the theoretical means, $\mu_i$, since, increasing the statistics, the bin
contents would approach the mean values with decreasing relative
uncertainty. This would be practical if we knew from the beginning which
specific model we want to test, but this procedure is not efficient if we want
to scan the parameter space, testing thousands or millions of models (points
in that space). So, identifying $u_i$ with $\mu_i$ is not justified unless
$u_i$ is large. The relation between them is given by
eq.(\ref{poiss_ui}). Since we are not sure about the values of $\mu_i$, we
cannot directly calculate the likelihood $\P(v_i)$ from
eq.(\ref{poiss_vi}). The best we can do is calculate $\P(v_i|u_i)$, i.e. the
probability of getting the experimental data, $v_i$, under the assumption that
the model is the true one (null-hypothesis), given that the simulation has produced $u_i$,
\bea
\label{nosys1}
P(v_i|u_i)=\int \prod_{i=1}^{K}\ d\mu_i \P(v_i|\mu_i)\P(\mu_i|u_i) .
\eea
Here $\P(v_i|\mu_i)$ is given by the Poisson distribution (\ref{poiss_vi}) and $\P(\mu_i|u_i)$ denotes the probability that the theoretical means are $\mu_i$, given that the simulation has produced the $u_i$--histogram. $\P(\mu_i|u_i)$ is not known, we must infer it using the Bayes theorem, 
\bea
\label{nosys2}
P(\mu_i|u_i)=\frac{\P(u_i|\mu_i)\P(\mu_i)}{\int d\mu_i \P(u_i|\mu_i)\P(\mu_i)} ,
\eea
where $\P(u_i|\mu_i)$ is the probability for each individual bin, given by the Poisson distribution (\ref{poiss_ui}), and $\P(\mu_i)$ is the prior for $\mu_i$. Since $\P(u_i|\mu_i)$ is peaked around $u_i=\mu_i$, the dependence on the prior, $\P(\mu_i)$, is small, but nevertheless it is there. The simplest procedure here is to take a flat prior for $\P(\mu_i)$. Then the $\P(\mu_i)$ cancels in the numerator and the denominator of eq.(\ref{nosys2}) (the latter becomes simply 1), and we can identify
\bea
\label{identify}
\P(\mu_i|u_i)\equiv\P(u_i|\mu_i) .
\eea
Now eq.(\ref{nosys1}) reads
\bea
\label{nosys3}
P(v_i|u_i)=\int \prod_{i=1}^{K}\ d\mu_i \frac{\mu_i^{v_i}}{v_i!}e^{-\mu_i} \frac{\mu_i^{u_i}}{u_i!}e^{-\mu_i}=\prod_{i=1}^{K} \frac{(u_i+v_i)!}{u_i!\ v_i!}\ \ 2^{-1-u_i-v_i} .
\eea
This formula represents our best estimate the likelihood, although is only valid when the non-statistical sources of uncertainty, both in the experimental and in the theoretical side, are ignored (they are incorporated in the next section). Note that expression (\ref{nosys3}) 
avoids the problem of the empty bins in the theoretical simulation. In other words, if one simply identified $\mu_i=u_i$, then the presence of an empty bin ($u_i=0$) would make the whole likelihood - eq.(\ref{poiss_vi})- vanishing. Therefore the $\P(\mu_i|u_i)$ piece in the calculation of the likelihood, eq.(\ref{nosys1}), is important, at least for bins with low statistics. 

\subsection{Separation of normalization and shape tests}

Suppose for a moment we could calculate all $\mu_i$ with great accuracy and that we keep ignoring other sources of uncertainties different from the statistical ones. Then, the likelihood is simply given by the Poisson distribution $\P(v_i)$, as given by eq.(\ref{poiss_vi}).

Now, it is interesting that that expression can be separated in a test for the
global normalization (the total number of events) and a test for the
shape. Namely
\bea
\label{separate}
P(v_i)=\prod_{i=1}^{K}\ \frac{\mu_i^{v_i}}{v_i!}e^{-\mu_i}=\P({\rm norm})
\times \P({\rm shape}) ,
\eea
where
\bea
\label{norm_shape}
\P({\rm norm})&=&\frac{\mu^{v}}{v!}e^{-\mu} \hspace{1.5cm} {\rm with}\;\; \mu=\sum_i \mu_i ,
\nonumber\\
\P({\rm shape})&=&\prod_{i=1}^{K}\ \frac{\mu_i^{v_i}}{\mu^{v_i}}\frac{(v!)^{1/K}}{v_i!}=
V\ \prod_{i=1}^{K}\ (\mu_i\frac{v}{\mu})^{v_i} ,
\eea
with
\bea
\label{V}
V=\frac{v!}{v^v}\prod_{i=1}^{K}\frac{1}{v_i!}\ =\ {\rm const.}
\eea

Notice that both $\P({\rm norm})$, $\P({\rm shape})$ are proportional to terms given by Poisson distributions. In particular, $\P({\rm shape})$ is proportional to a Poisson distribution, where the means $\mu_i$ are re-normalized so that they would fit perfectly the total number of events:
\bea
\label{shape}
\P({\rm shape})&=&V \prod_{i=1}^{K} (\mu_i\frac{v}{\mu})^{v_i} \nonumber \\
&=&(e^{v}\prod_{i=1}^{K}  v_i!)\ V\ \prod_{i=1}^{K} 
\frac{(\mu_i\frac{v}{\mu})^{v_i}}{v_i!}e^{-\mu_i\frac{v}{\mu}} \nonumber \\
&=& 
\frac{v!e^v}{v^v} \prod_{i=1}^{K} \frac{(\mu_i\frac{v}{\mu})^{v_i}}{v_i!}e^{-\mu_i\frac{v}{\mu}} .
\eea
This expression is really independent of the global normalization. i.e. if we make $\mu_i\rightarrow a\mu_i$, then $\P({\rm shape})$ remains the same. This also tells us that if we fix the shape of a simulated histogram and allow to change its global normalization (i.e. we allow $\mu_i\rightarrow a\mu_i$), the total probability (\ref{separate}) is always maximal when the global mean, $\mu$, coincides with the total number of events, $v$.
 
The interesting thing about separating normalization and shape tests is that
one can treat the extra sources of systematic uncertainty for both in a separate way, as will become clear in the next section. For instance, one may consider that the amount of 
systematic uncertainty in the global normalization is larger than in the shape, and hence it is useful to separate the two tests.

\section{Incorporating other sources of uncertainty. Systematic errors}
\label{sec:errors}

\subsection{General strategy}

There are several sources of uncertainty in the comparison of the experimental
data with the theoretical predictions. First, there is the statistical
uncertainty, associated to the Poisson distributions, which has been the
subject of the previous section. Besides there are additional sources of
systematic uncertainty, both in the experimental side (the resolution
and the scale of jets and missing transverse momentum, b-tagging, pile-up,
etc.) and in the theoretical one (K-factors, parton distribution functions, etc.). However, for practical purposes, we can treat the experimental data as if they were free from systematic errors and ``absorb" {\em all} the experimental systematic uncertainty in the theoretical side. 

We will call $\mu_i^{th}$ the means that, in the simulation process, have produced the theoretical ($u_i$) histogram. Now, due to the systematic uncertainty, we {\em cannot} identify them directly with the ``true means", $\mu_i$, which are the real ones associated with the model under consideration, and thus the ones that, supposedly, have ``produced" the experimental histogram ($v_i$) under the null-hypothesis. The relation between them can be expressed as
\bea
\label{transfer}
\mu_i(M)\ =\ F(M_i)\ \muth_i ,
\eea
where $F(M)$ is some ``transfer function" on the effective mass ($M$) that encodes all (experimental and theoretical) systematic uncertainties. This function can depend on a number of unknown parameters, though we know it cannot be completely arbitrary (below we give an ansatz for $F(M)$).

Now, in analogy with eq.(\ref{nosys1}), the best estimate for the likelihood is
\bea
\label{complete}
P(v_i|u_i)=\int DF \int D\muth_i\ \P(v_i|\mu_i)\ \P(\muth_i|u_i)\ \P(\mu_i|\muth_i) ,
\eea
where $\P(\mu_i|\muth_i)\equiv \P(F)$ is still to be guessed, and the 
integration measure $DF$ is written in a symbolic form. The first two factors in the integrand are statistical probabilities, as in (\ref{nosys1}). The third factor contains the systematic uncertainty (if we decided to ignore it, then we would simply take $\P(\mu_i|\muth_i) \equiv \delta(F-1)$).

We can write explicit expressions for the three factors in eq.(\ref{complete}). The first factor,$\P(v_i|\mu_i)$, is given by the Poisson distribution (\ref{poiss_vi}). Regarding the second factor, recall that (taking a flat prior for $\muth_i$) we can identify 
\bea
\label{Pumuth}
\P(\muth_i|u_i)\ \sim \P(u_i|\muth_i)\ =\ \prod_{i=1}^{K} \frac{(\muth_i)^{u_i}}{u_i!}e^{-\muth_i} .
\eea
Finally, we have to make ansatz for the $F$ function and its probability, $\P(\mu_i|\muth_i)\equiv \P(F)$. Since it is convenient to separate the uncertainties associated to the global normalization and to the shape, we express eq.(\ref{transfer}) as
\bea
\label{mumuth}
\mu_i \ =\ F(M_i)\ \muth_i\ =\ f\ g_i\ \muth_i .   
\eea
Here $f$ and $g_i$ carry the uncertainty in the global normalization and in the shape, respectively. With this definition, $g_i$ obey the relation
\bea
\label{condg}
\sum_i g_i\ \muth_i\ =\ \sum_i \muth_i \ \equiv\ \muth ,
\eea
i.e. the $g_i$ parametrize systematic errors that modify the shape of the histogram without changing the total number of events. The situation $f=g_i=1$ corresponds to the absence of systematic errors, but we have to assign a non-vanishing probability to the possibility that $f, g_i$ depart from that ideal situation. Thus we write 
\bea
\label{Pfg}
\P(\mu_i|\muth_i)\equiv \P(f, g_i)= \P(f)\ \P(g) . 
\eea
For the moment we do not write a concrete ansatz for $\P(f)$, $\P(g)$ (this is postponed to the next subsection). So, the likelihood (\ref{complete}) is given by
\bea
\label{Pviui}
\P(v_i|u_i)=\int D\muth_i \int Df Dg\ \left(\prod_{i=1}^{K}   
\frac{\mu_i^{v_i}}{v_i!}e^{-\mu_i}\right) \left(\prod_{i=1}^{K} \frac{(\muth_i)^{u_i}}{u_i!}e^{-\muth_i}\right) \P(f)\P(g) ,
\eea
In this expression $Df$, $Dg$ are symbolic ways to express integration over all the possibilities for $f$, $g_i$.

\subsection{Ans\"atze for the transfer functions}

In eq.(\ref{mumuth}) we have written the ``transfer" function, $F$, that encodes the systematic uncertainty, as
\bea
\label{Ffg}
F(M_i) \ =\ f\ g_i ,     
\eea
but so far we have not established on which parameters the $F$-function --and
thus the quantities $f$, $g_i$-- depend. A simple and handy choice for
practical purposes is to take the very values of $\{f,g_i\}$ as those
independent parameters. Alternatively, since systematic errors must depend on
$M$ in a smooth way, we could parametrize $F(M)$ as a smooth function,
e.g. $F\sim f\ \sum_\alpha a_\alpha P_\alpha$, where $P_\alpha$ are $\sim$ Legendre Polynomials and the summation contains just a few terms. Then, the $F$ function would be defined by the $a_\alpha$ coefficients (together with the global normalization factor, $f$). This would be sensible, but it leads to very cumbersome expressions, difficult to handle. On the other hand, since in practice $v_i$ and $u_i$ are both quite smooth (apart from statistical noise), only sets of values of
$F(M_i)$ that vary smoothly with $M$ can lead to a simultaneous fit of both histograms. In other words, chaotic values of $F(M_i)$ (or, equivalently, $g_i$) will be strongly penalised by the $\P(v_i|\mu_i)$ piece (first factor in eq.(\ref{Pviui})). So, even if those eccentric choices for $g_i$ are not specially penalised by $\P(g)$,
they are by other factors in the likelihood and become irrelevant. In consequence, choosing $\{f,g_i\}$ as independent parameters is a reasonable option

Concerning the integration measures, we could simply take $Df=df$, $Dg=\prod_i dg_i$. However, since $\{f,g_i\}$ are defined as multiplicative factors in eq.(\ref{mumuth}), it seems much more sensible to use their magnitudes as the actual unknowns. This is equivalent to choose $\{\ln f, \ln g_i\}$ as the independent parameters. Then,
\bea
\label{DfDg}
Df\equiv \frac{1}{f}\ df,\;\;\;\; Dg\equiv \prod_{i=1}^K\frac{1}{g_i}\ dg_i . 
\eea
Of course, since $\{f,g_i\}$ are never far from 1, it does not make a big difference to use $\{f,g_i\}$ or $\{\ln f, \ln g_i\}$, but it can be checked that the second option leads to a more stable and satisfactory test. Note that, in principle, the $g_i$ variables are subject to condition (\ref{condg}), so there are in fact $K-1$ independent $g_i$ variables. However, for the moment we have ignored such complication in writing (\ref{DfDg}).

Finally, concerning the probabilities $\P(f)$, $\P(g)$, we can take them as gaussians centered around $f=g_i=1$. The argument of these gaussians must be essentially the ``squared-distance" of $\{f,g_i\}$ to their central values, i.e. $\P(f)\sim \exp\{-\frac{1}{2}(f-1)^2\}$ and $\P(g)\sim \exp\{-\frac{1}{2}\int dM (g(M)-1)^2\}\ \sim \exp\{-\frac{1}{2}\sum_i (g_i-1)^2\}$. A nice fact here is that $\P(g)$ appears naturally factorized as $\prod_i \P(g_i)$, which is very convenient for analytical manipulations.

A suitable (and equivalent at first order), way to express these ans\"atze is by using the logarithmic variables, $\{\ln f, \ln g_i\}$:
\bea
\label{Pf2}
\P(f)=\frac{1}{\sqrt{2\pi}\Delta_f}e^{-\frac{1}{2}\left(\frac{\ln f}{\Delta_f}\right)^2} ,
\eea
\bea
\label{Pg2}
\P(g)\propto \frac{1}{\Delta_g^K}e^{-\frac{1}{2}\sum_i\left(\frac{\ln \ g_i}{\Delta_g}\right)^2} ,  
\eea
where the widths $\Delta_f$, $\Delta_g$ measure our degree of ignorance about the magnitude of $f$, $g_i$.
Note that the use of logarithmic variables allows to maintain the whole range of integration of the gaussians, $[-\infty, \infty]$, without artificial cuts to keep $\{f, g_i\}$ positive. 

In any case, we will go as far as possible in the analysis without specifying the precise ans\"atze for $\P(f)$, $\P(g)$.

\subsection{Separation of normalization and shape tests}

Coming back to our expression (\ref{Pviui}) for the likelihood, we note 
that the first factor of (\ref{Pviui}) may be decomposed, as in eqs.(\ref{separate})-(\ref{norm_shape}), into a factor for the global-normalization  and another for the shape:
\bea
\label{normdeltav}
\left(\prod_{i=1}^{K}   
\frac{\mu_i^{v_i}}{v_i!}e^{-\mu_i}\right)\ =\ \frac{\mu^{v}}{v!}e^{-\mu}\ V\ 
\prod_{i=1}^{K}\ (\mu_i\frac{v}{\mu})^{v_i} ,
\eea
where $V$ is given in eq.(\ref{V}). Since the {\em total} number of events is normally large the global-normalization factor can be approximated by a Dirac delta,
\bea
\label{decomp_v}
\frac{\mu^{v}}{v!}e^{-\mu}\ 
\simeq \delta(\mu-v)=\delta(f\muth-v)=\frac{1}{\muth}\delta(f-v/\muth) .
\eea
Analogously, the second factor of (\ref{Pviui}) can be written as
\bea
\label{decomp_u}
\prod_{i=1}^{K} \frac{(\muth_i)^{u_i}}{u_i!}e^{-\muth_i}\ \simeq\ \delta(\muth-u)\times
U\ \prod_{i=1}^{K}\ (\muth_i\frac{u}{\muth})^{u_i} ,
\eea
with
\bea
\label{U}
U=\frac{u!}{u^u}\prod_{i=1}^{K}\frac{1}{u_i!}\ =\ {\rm const.}
\eea
We can use the presence of these deltas to extract pieces of the integrand of eq.(\ref{Pviui}) outside the sign of integration. Hence,
\bea
\label{Pviui2}
\P(v_i|u_i) &\propto& \P(f=\frac{v}{u}) \nonumber \\
&\times& \int D\muth_i Dg \left(\prod_{i=1}^{K}   
\frac{(\frac{v}{u}g_i\muth_i)^{v_i}}{v_i!}e^{-\frac{v}{u}g_i\muth_i}\right) \left(\prod_{i=1}^{K} \frac{(\muth_i)^{u_i}}{u_i!}e^{-\muth_i}\right) \P(g) .
\eea
Note that we have made explicitely the integration in $\int Df=\int (1/f)df$, but not in $\int d\muth$. However the implicit presence of the $\delta(\muth-u)$ in the integrand, as expressed in eq.(\ref{decomp_u}), has allowed us to replace $\muth\rightarrow u$ in a consistent way.

Assuming in the previous expression that $Dg$ and $\P(g)$ are factorizable as
products of $K$ factors, like in eqs.(\ref{DfDg}), (\ref{Pg2}), makes much
easier the integration in practice. As mentioned, $g_i$ are subject to condition (\ref{condg}), so strictly speaking we only have $K-1$ independent $g_i$ variables and this factorization is not complete. In spite of this, assuming a complete factorization is a sensible and good approximation. The reason is the following. In eq.(\ref{Pviui2}) the Poisson distribution in the first factor of the integrand only departs appreciably from zero when 
$\sum g_i\muth_i\simeq u\simeq\muth$. This can be checked by doing again a decomposition of such distribution as in (\ref{separate}, \ref{norm_shape}) and noting that the global-normalization piece of the decomposition is essentially a Dirac delta, $\delta(\sum \frac{v}{u}g_i\muth_i-v)$. So, even assuming that $Dg$ and $\P(g)$ are factorizable, and thus integrating over sets of $\{g_1, g_2, ...g_K\}$ which do not respect (\ref{condg}), the Poisson distribution in the first factor of the integrand causes that only those sets that obey condition (\ref{condg}) will contribute appreciably to the integral. Alternatively one could understand this procedure considering that in the expression (\ref{mumuth}), the $g_i$ variables that encode $M_i-$dependent systematic errors can also distort the total number of events. This is in fact a quite realistic situation. Then, the relation (\ref{condg}) is not to be imposed and $Dg$ and $\P(g)$ become factorizable as products of $K$ factors (strictly speaking). The trouble is that the previous separation between normalization and shape cannot be done exactly. But, if the $g_i$ only amount to slight distorsions of the total normalization (in other words, $\P(g)$ penalizes much more severely the variation in the normalization than $\P(f)$) the separation (\ref{decomp_u}) is a good approximation and (\ref{Pviui2}) is valid.

Now, taking profit of the factorization of $Dg$ and $\P(g)$ we can make explicitely the integration in the $\muth_i$ variables, with no need of specifying the ans\"atze for $\P(f)$, $\P(g)$:
\bea
\label{Pviui3}
\P(v_i|u_i)\ &\propto &\  \P(f=\frac{v}{u})
\nonumber\\
&\times& \prod_{i=1}^{K}\left( \frac{(u_i+v_i)!}{u_i! v_i!}\ \int dg_i \frac{1}{g_i}
\left(\frac{v}{u}g_i\right)^{v_i}\left(1+\frac{v}{u}g_i\right)^{-1-u_i-v_i}
\P(g_i)\right) .
\eea
In this expression, the factor of the first line, $\P(f=\frac{v}{u})$, carries
the test for the global normalization: it is only sensitive to the mismatch
between the experimental total number of events, $v$, and the theoretical one,
$u$. The remaining factor (second line) corresponds to the test of the
shape. It is interesting to check that indeed, for given $u,v$, this expression has a maximum at $u_i = (u/v) v_i$.

Eq.(\ref{Pviui3}) represents our final expression to evaluate the likelihood of a simulated histogram, $u_i$, confronted to the experimental one, $v_i$. (A modified version is given in eq.(\ref{Pviui3loggorro}) of Appendix A to incorporate the fact that the luminosity of the simulated histogram may be different from that of the experimental one.) This expression amounts to realize $K$ integrals, which can be done numerically at low cost in computing time, even if one needs to probe thousands or millions of histograms, corresponding to points in the parameter space of a theoretical scenario. All this is illustrated in the next section.

\section{Application to the CMSSM}
\label{sec:cmssm}

\subsection{Set up}

In this section we apply the previous histogram-comparison techniques to the study of the Minimal Supersymmetric Standard Model (MSSM) \cite{Martin:1997ns}. More precisely, we will consider the somehow standard framework, often called CMSSM or MSUGRA, in which the soft parameters are assumed universal at a high scale ($M_X$), where the supersymmetry (SUSY) breaking is transmitted to the observable sector; as happens e.g. in the gravity-mediated SUSY breaking scenario. Hence, our parameter-space is defined by the following parameters:
\bea
\label{MSSMparameters}
\{\theta_i\}\ =\ \{m,M_{1/2},A,B,\mu,s\} \ .
\eea
Here $m$, $M_{1/2}$ and $A$ are the universal scalar mass, gaugino mass and trilinear scalar coupling; $B$ is the bilinear scalar coupling; $\mu$ is the usual Higgs mass term in the superpotential; and $s$ stands for the SM-like parameters of the MSSM. The latter include the $SU(3)\times SU(2)\times U(1)_Y$ gauge couplings, $g_3,g,g'$, and the Yukawa couplings, which in turn determine the fermion masses and mixing angles. All the initial parameters are defined at the $M_X$ scale.

The goal is to scan the CMSSM parameter space, determining the most probable region in it, given the available (present or future) experimental (mainly LHC) data. To show the power of the histogram-comparison technique, we will simulate LHC data assuming that nature lives in a standard benchmark SUSY model. This simulation will be considered as our (mock) experimental data. Then we will scan the CMSSM parameter space using Bayesian techniques to find out the most probable region of parameters, showing to which extent the histogram-comparison between the mock data and the theoretical prediction is capable to determine the ``true" model.

As mentioned in sect. \ref{sec:histograms}, in a Bayesian analysis the most important quantity is the {\em posterior} probability density function (pdf) in the parameter space, which is given by the fundamental Bayes' relation
\bea
\label{pdf}
p(s, m, M_{1/2}, A, B, \mu |{\rm data})\ \propto\ p({\rm data}|s, m, M_{1/2}, A, B, \mu)\ 
p(s, m, M_{1/2}, A, B, \mu ) ,
\eea
where the first factor in the right hand side is the {\em likelihood} (probability of measuring the observed data assuming that the point in the parameter space is the true model) and the second one is the {\em prior} (probability assigned to that point before knowing the experimental data). Next, we discuss the precise form of the pdf (\ref{pdf})  for the problem at hand.

First of all, it should be noticed that very often in statistical problems not
all the parameters that define the system are of the same interest. The usual
technique to eliminate the less interesting ones from the problem is simply
marginalizing them, i.e. integrating the pdf (\ref{pdf}) in those variables
(for a review see ref. \cite{Berger:1999}). This is the standard procedure to
deal with the {\em nuisance parameters} $\{s\}$. Besides, for the purposes of
scanning the CMSSM parameter space it is convenient to trade some of the
initial parameters (\ref{MSSMparameters}) by others with more direct
phenomenological significance. We follow here the approach expounded in detail
in refs. \cite{Cabrera:2008tj, Cabrera:2009dm}. As usual, the value of $\mu$
can be traded by the value of $M_Z$ using the minimization conditions of the
Higgs potential. The Yukawa couplings can be traded by the physical fermion
masses; in particular the top Yukawa coupling, $y_t$, can be traded by the top mass, $m_t$. Finally, it is
  highly advantageous to trade the initial $B-$parameter by the derived $\tan\beta\equiv \langle H_u\rangle/\langle H_d\rangle$ parameter, where $H_u, H_d$ are the Higgs doublets of the MSSM\footnote{This change of variables still leaves the sign of $\mu$ undetermined. For simplicity we have assumed a positive $\mu$ in the rest of the paper.}.

Consequently, to write the pdf in the new variables one should compute the Jacobian, $J$, of the transformation
\bea
\label{change_3}
\{\mu,y_t,B\}\ \rightarrow\  \{M_Z,m_t,\tan\beta\} .
\eea
On the other hand, the $\mu$-parameter (now traded by $M_Z$) can be easily marginalized, together with the nuisance parameters, taking profit of the high precision of our knowledge of $M_Z^{exp}$. Consequently, the final expression for the posterior (\ref{pdf}) in the new variables is
\bea
\label{pdf2}
p(m, M_{1/2}, A, \tan\beta |{\rm data})\ &\propto&\ J|_{\mu=\mu_Z}\ p({\rm data}|m, M_{1/2}, A, \tan\beta)
\nonumber\\
&\times& p(m, M, A, B, \mu=\mu_Z) ,
\eea
where $J$ is the Jacobian of the transformation
(\ref{change_3}) and $\mu_Z$ is the value of $\mu$ that reproduces $M_Z^{\rm exp}$ for the given values of $\{m, M, A, \tan\beta\}$. Let us discuss in order the three factors of the r.h.s. of (\ref{pdf2}).

The Jacobian factor, $J$, has to be evaluated using the (radiative) electroweak breaking conditions of the CMSSM. For the numerical analysis we have computed $J$ using the \texttt{SoftSusy} code \cite{softsusy} which implements the full one-loop contributions and leading two-loop terms to the tadpoles for the those conditions, with parameters running at two-loops. This essentially corresponds to the next-to-leading log approximation. A quite accurate analytical expression of $J$, corresponding to the leading log approximation, reads \cite{Cabrera:2008tj} 
\bea
\label{J_anal}
J|_{\mu=\mu_Z}\ \propto \   \left[\frac{E}{R_\mu^2}\right]\ 
\frac{y}{y_{\rm low}} \frac{\tan\beta^2-1}{\tan\beta(1+\tan^2\beta)} 
\frac{B_{\rm low}}{\mu_Z} . 
\eea
Here $y$ denotes the top Yukawa coupling and the ``low'' subscript indicates that the quantity is evaluated at low scale (say $Q_{\rm low}\equiv$ typical supersymmetric mass). $R_\mu$ and $E$ are RG quantities, involved in the one-loop running of $\mu$ and $y$:
\bea
\label{muBLH}
\mu_{\rm low}= R_\mu\ \mu,\;\;\;\; y_{\rm low}\simeq \frac{y E(Q_{\rm low})}{1+6yF(Q_{\rm low})} ,
\eea
where $Q$ is the renormalization scale and $F = \int_{Q_{\rm high}}^{Q_{\rm low}} E \ln Q$. 
$R_\mu$ and $E$ are definite functions that depend just on the top Yukawa
coupling and the gauge couplings, respectively, \cite{Martin:1997ns,Ibanez:1983di}. 

The important point about the Jacobian is that it does not represent any subjective prior on the parameters. Such subjectivity is still contained in the prior factor that stands in the second line of eq.(\ref{pdf2}). The Jacobian is simply a consequence of scanning the MSSM parameter space in some variables, which are not the initial ones, but derived quantities. Another important point is that $J$ automatically incorporates a penalization of the regions that require fine-tuning in order to reproduce the correct electroweak scale (typically regions of large soft parameters), as well as a penalization of large $\tan\beta$, reflecting also the fine-tuning needed to implement such possibility. A more detailed discussion of these issues can be found in refs. \cite{Cabrera:2008tj, Cabrera:2009dm}.

\begin{center}
\begin{figure}[t]
\label{meff}
\includegraphics[angle=0,width=0.95\linewidth]{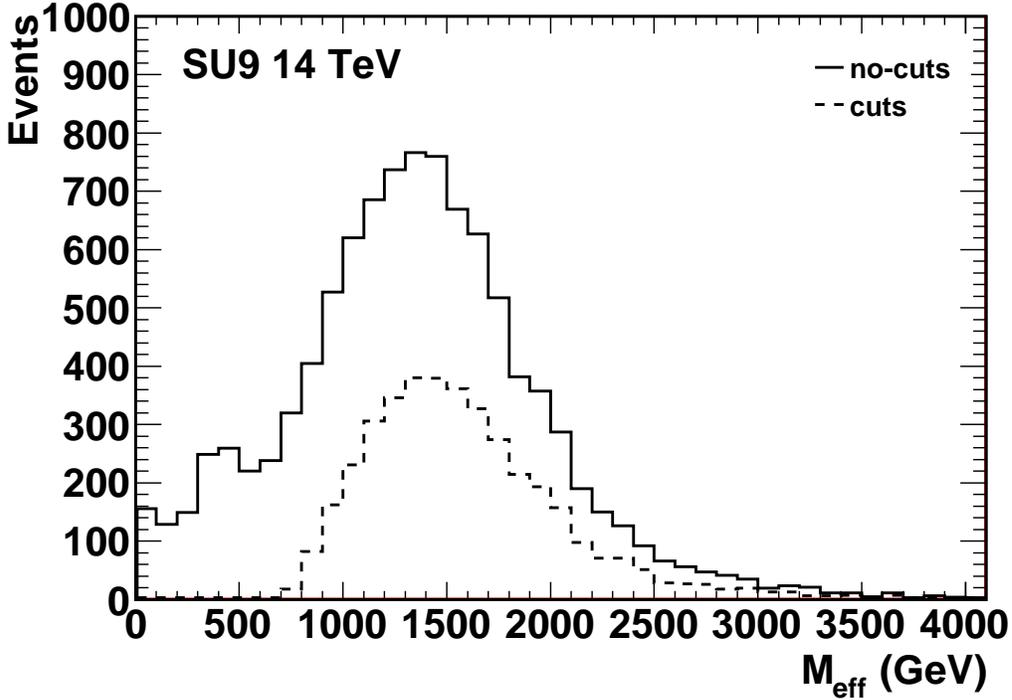} \hspace{1.2cm}
\caption[text]{Effective mass distribution expected for the SUSY SU9 model without and with cuts applied as given in subsect. 4.1.}
\label{fig:histo}
\end{figure}
\end{center}

Let us discuss now the second factor of the r.h.s. of eq.(\ref{pdf2}), i.e. the likelihood. This consists of a product of likelihoods corresponding to the experimental observables used in the analysis. For the present one, we just consider (besides the value of $M_Z^{\rm exp}$) the experimental bounds on the masses of supersymmetric particles and the lightest Higgs boson (see ref.\cite{deAustri:2006pe} for details), and the mock LHC data of multijet events plus missing transverse momentum, which is the main focus of this section\footnote{Most of the electroweak precision tests are currently being surpassed by LHC data. Other observables, like $b\rightarrow s, \gamma$ would have a moderate impact in the analysis, but we want to focus on the impact of the LHC, which is already the dominant part of the likelihood. For the same reason we have not included the somewhat controversial $g-2$ data or Dark Matter constraints.}. More precisely, for the sake of the simulated LHC data we work under the hypothesis that nature lies in the so-called SU9 benchmark point, defined in ref. \cite{Aad:2009wy}. This is specified by the following values of the CMSSM parameters:
\bea
\label{SU9}
m=300\ {\rm GeV},\;\;\;M_{1/2}=425\ {\rm GeV},\;\;\;A=20\ {\rm GeV},\;\;\;\tan\beta=20,\;\;\;\mu>0 .
\eea
The corresponding values of the squark mass (first two generations) and the gluino mass are $m_{\tilde q}=920$ GeV, $M_{\tilde g}=994$ GeV. This point of the CMSSM parameter space is on the verge of being excluded by the last analyses by ATLAS and CMS \cite{atlas:last, cms:last}. Of course, assuming the SU9 point is just an example to show the histogram-comparison technique at work, combined with Bayesian analysis. The LHC simulation has been performed using Pythia version 6.419 \cite{Sjostrand:2006za} with events generated at $E_{CM}=14$ TeV, and selecting those satisfying the following cuts\footnote{We have followed the strategy given in sect. 13.5 of ref. \cite{Ball:2007zza}}:
\begin{itemize}
\item Three or more jets with $p_T>30$ GeV and $|\eta|<3.0$. The hardest with $p_T>180$ GeV and $|\eta|<1.7$, the second with  $p_T>110$ GeV .
\item $p_T^{\rm miss}>200$ GeV .
\item $\Delta\phi_1>0.3$,\ \ $\Delta\phi_2>0.3$,\ \ $\Delta\phi_3>0.3$ .
\item $\sqrt{\Delta\phi_2^2 + (\pi-\Delta\phi_1)^2}>0.5$,  $\sqrt{\Delta\phi_1^2 + (\pi-\Delta\phi_2)^2}>0.5$ and $\Delta\phi_2>\pi/9$ .
\item $H_T = \sum_{i=2}p^i_T + p_T^{\rm miss} >500$ GeV .
\end{itemize} 
where $\Delta\phi_i \equiv \Delta\phi({\rm jet}_i-p_T^{\rm miss})$. 
Concerning the luminosity we have considered $10^4$ supersymmetric events,
upon which we impose the previous cuts. Since the total cross section for SUSY
production in the SU9 model is $2.4$ pb, this corresponds to a luminosity of
about $4.2$ fb$^{-1}$. The histogram of number of events as a function of the
effective mass, $M_{\rm eff}$, is shown in Fig. \ref{fig:histo}, where
only the SUSY events have been displayed. For each event, $M_{\rm eff}$ is defined as \cite{Aad:2009wy}
\bea
\label{Meff}
M_{\rm eff}\  = \   \sum_{j}|p_T^j| + p_T^{\rm miss} .
\eea
where $j$ runs over all jets satisfying the previous cuts. Note that the latter
effectively imply a lower bound $M_{\rm eff}\geq 680$ GeV for the events considered.

\vspace{-1.5cm}
\noindent
Using the notation of sects. \ref{sec:histograms}, \ref{sec:errors}, the bin contents of the ``experimental" histogram of Fig. \ref{fig:histo} are the $v_i$ quantities. For each point scanned in the parameter space we compute a simulated histogram (the $u_i$ quantities in sects. \ref{sec:histograms}, \ref{sec:errors}), and then evaluate the likelihood through eq. (\ref{Pviui3}).
In order to use that expression, we have to specify the $\P(f)$, $\P(g)$ functions, which encode
the systematic uncertainty assigned to the total number of events and the shape of the histograms respectively.
In our case, we have used for them the gaussian profiles (\ref{Pf2}), (\ref{Pg2}). The width of the first, $\Delta_f$, reflects the uncertainty in the total number of events due to (mainly) theoretical uncertainties associated with the K-factors and the parton distribution functions. We have been pretty conservative, assuming $\Delta_f=0.5$; in other words we accept that a factor 2 or 1/2 in the total number of events is plausible\footnote{In the present scan we are using a tree-level Pythia simulation of the parton events; which implies an important uncertainty about the K-factors. In a 1-loop-refined simulation this uncertainty could be assumed smaller.}. On the other hand, from eq.(\ref{Pg2}), $\Delta_g$ goes as the typical systematic uncertainty in the shape times $\sqrt{K}$. For instance if, {\em once} the uncertainty affecting the global normalization is extracted, one estimates that there remains a systematic uncertainty in the shape which is of order 10\% for every bin, then one has to use $\Delta_g\simeq 0.1\sqrt{K}$ to reproduce that uncertainty at 1$\sigma$. In our case the total number of bins is $K=10$, so we estimate $\Delta_g=0.2$ as a reasonable choice.

To conclude our discussion of the likelihood, let us note that the total cross section varies from point to point in the CMSSM parameter space. This implies that considering $10^4$ initial supersymmetric events does not correspond to the same luminosity for each point scanned in the parameter space. But of course, the comparison of histograms must be realized under the same conditions of luminosity. This feature can be easily incorporated into the histogram-comparison technique discussed in sect. \ref{sec:errors}, leading to a slight and straightforward correction in the expression (\ref{Pviui3}) for the likelihood, which now becomes eq.(\ref{Pviui3loggorro}) of Appendix A (see that appendix for more details). And that is the expression that we have finally used to compute the likelihood when scanning the CMSSM parameter space.

We end up this subsection with a brief discussion of the third factor of the r.h.s. of eq.(\ref{pdf2}), i.e. the prior in the initial parameters. Admittedly, this is the less objective part of the statistical analysis, but one cannot simply ignore the prior. This would be equivalent to take a flat prior in the initial parameters, a choice which is as arbitrary as any other, unless one can give some argument of plausibility for it. Besides, in order to perform the marginalizations one should specify the ranges where the parameters live\footnote{Fortunately, in our case this point is irrelevant thanks to the Jacobian factor, $J$. As mentioned above, $J$ automatically incorporates a fine-tuning penalization of the high-energy region of the parameters, which thus becomes irrelevant. For more details see \cite{Cabrera:2009dm}.}. The dependence on the prior is actually a measure of the dependence of the results of the statistical analysis on a priori assumptions or prejudices. Note here that when the likelihood factor is very sharp, i.e. it distinctly selects a narrow region in the parameter space, the prior factor becomes irrelevant, since the posterior can only be sizeable where the likelihood is. But unfortunately we are not in that ideal situation yet, so there {\em exists} a dependence on the prior. Hence, the most conservative attitude is to use two different, though still reasonable, priors, and then compare the results. This gives a fair measure of the prior-dependence.

For that matter we have considered two somehow standard types of prior: flat and logarithmic. In a flat (logarithmic) prior one assumes that, in principle, the typical size (order of magnitude) of the soft terms can be anything, say from 10 GeV up to $M_X$, with equal probability. In our opinion, a logarithmic prior is probably the most reasonable option, since it amounts to consider all the possible magnitudes of the SUSY breaking in the observable sector on the same foot (this occurs e.g. in conventional SUSY breaking by gaugino condensation in a hidden sector). However, we will consider log and flat priors at the same level throughout the paper, in order to compare the results and thus evaluate the prior-dependence. The precise forms of these priors can be found in sect. 2  of ref. \cite {Cabrera:2009dm}, together with a detailed discussion.

\subsection{Results}

We have computed the distribution of the posterior (\ref{pdf2}) in the CMSSM parameter space using a modified version of the public \texttt{SuperBayeS} package
\cite{SuperBayeS}\footnote{For this paper, the public \texttt{SuperBayeS} code has been modified to interface with Pythia 6.419 \cite{Sjostrand:2006za}.}
 adopting MultiNest v2.8 \cite{Feroz:2007kg,Feroz:2008xx} as a scanning
 algorithm. We use as running parameters a number of live points $n_\text{live} = 2000$ and a tolerance parameter $\text{tol} = 1$. Our final inferences for each of the log and flat priors are obtained from chains generated with approximately $10^5$ likelihood evaluations. 

We have also included in our likelihood the limits on the lightest Higgs and 
SUSY masses provided by LEP\footnote{Recent LHC bound on the Higgs mass are
  still irrelevant to constrain the MSSM parameter space, though this
  situation will change soon \cite{Cabrera:2011bi}} and Tevatron . For details on the implementation 
see ref. \cite{deAustri:2006pe}. 

For the marginalization procedure we have used $[0, M_X]$ as the range for $m$, $M_{1/2}$ and $|A|$. Besides, we have used $2 < \tan\beta < 62$. See a detailed discussion about this in sect. 2 of ref. \cite{Cabrera:2009dm}.

In order to show the potential of the histogram-comparison technique, we have performed the analysis twice: switching off and on the shape test. 

\subsubsection*{Test for the total number of events}

First we compute the (LHC-part of the) likelihood associated with a particular point in the CMSSM parameter space by comparing the prediction for the total number of supersymmetric events, satisfying the cuts specified in the previous subsection, with the ``experimental" result for that number (i.e. after subtracting the SM background). 

This means that for each point examined we compute an histogram of events, $u_i$, but we just compare the total number, $u=\sum u_i$, with the experimental one, $v=\sum v_i$, using the first factor of eq.(\ref{Pviui3}). More precisely, in order to incorporate the fact that the effective luminosity used in the simulation may change from point to point, we have actually used the --slightly modified-- first factor of eq.(\ref{Pviui3loggorro}),
\bea
\label{Norm}
{\rm LHC-likelihood}\ \propto \P(f=\frac{v}{Lu}) \ ,
\eea
where $L$ is the quotient of the experimental luminosity and the luminosity of the simulation. We recall that the $\P(f)$ function carries all the uncertainties affecting the total number of events, except the purely statistical ones (which are subdominant when that number is large), and is given by the gaussian (\ref{Pf2}) with $\Delta_f=0.5$; as explained in the previous subsection.

\begin{center}
\begin{figure}[t]
\hspace{-0.3cm}
\includegraphics[width=0.53\linewidth]{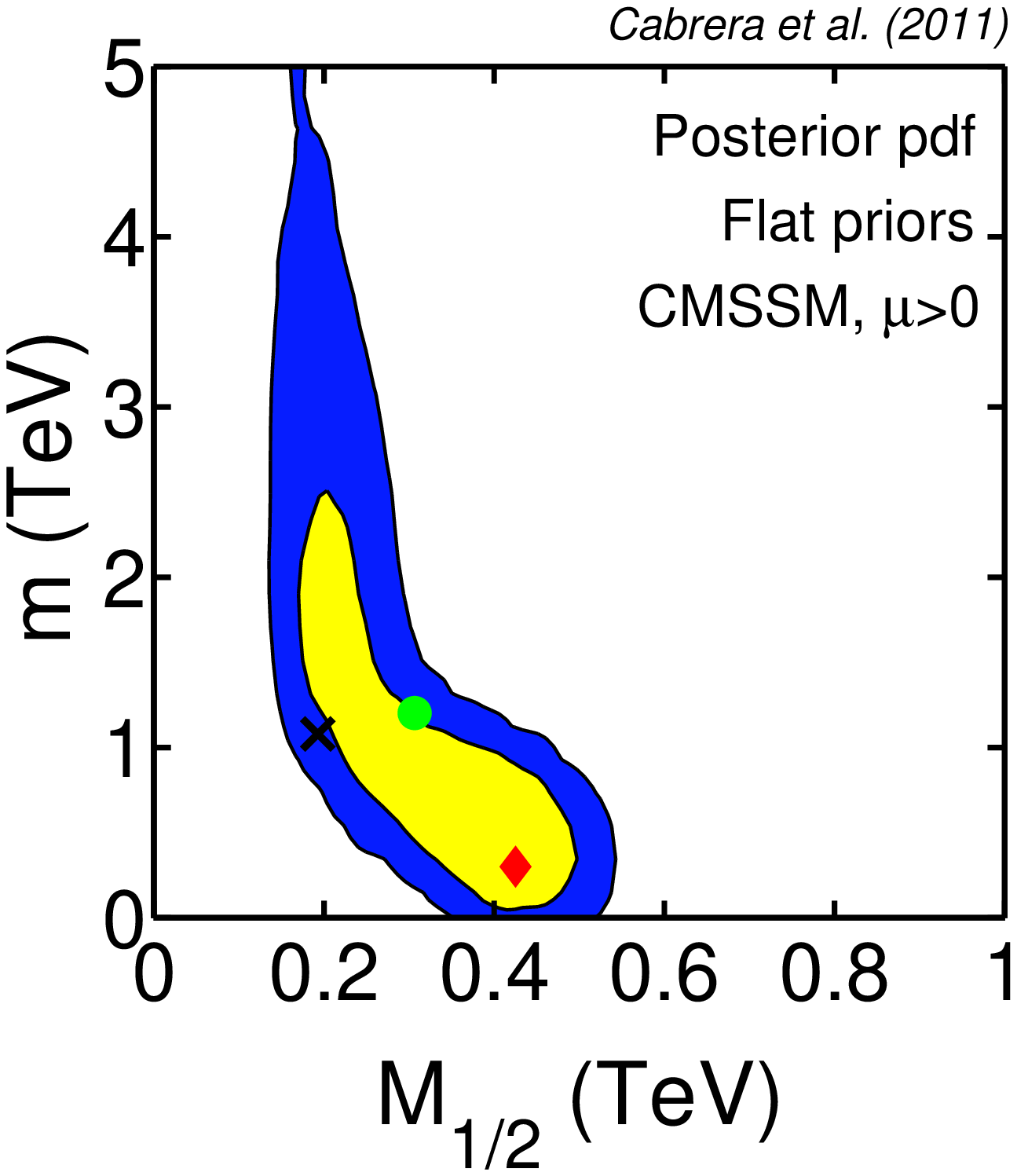} \hspace{-0.5cm}
\includegraphics[width=0.53\linewidth]{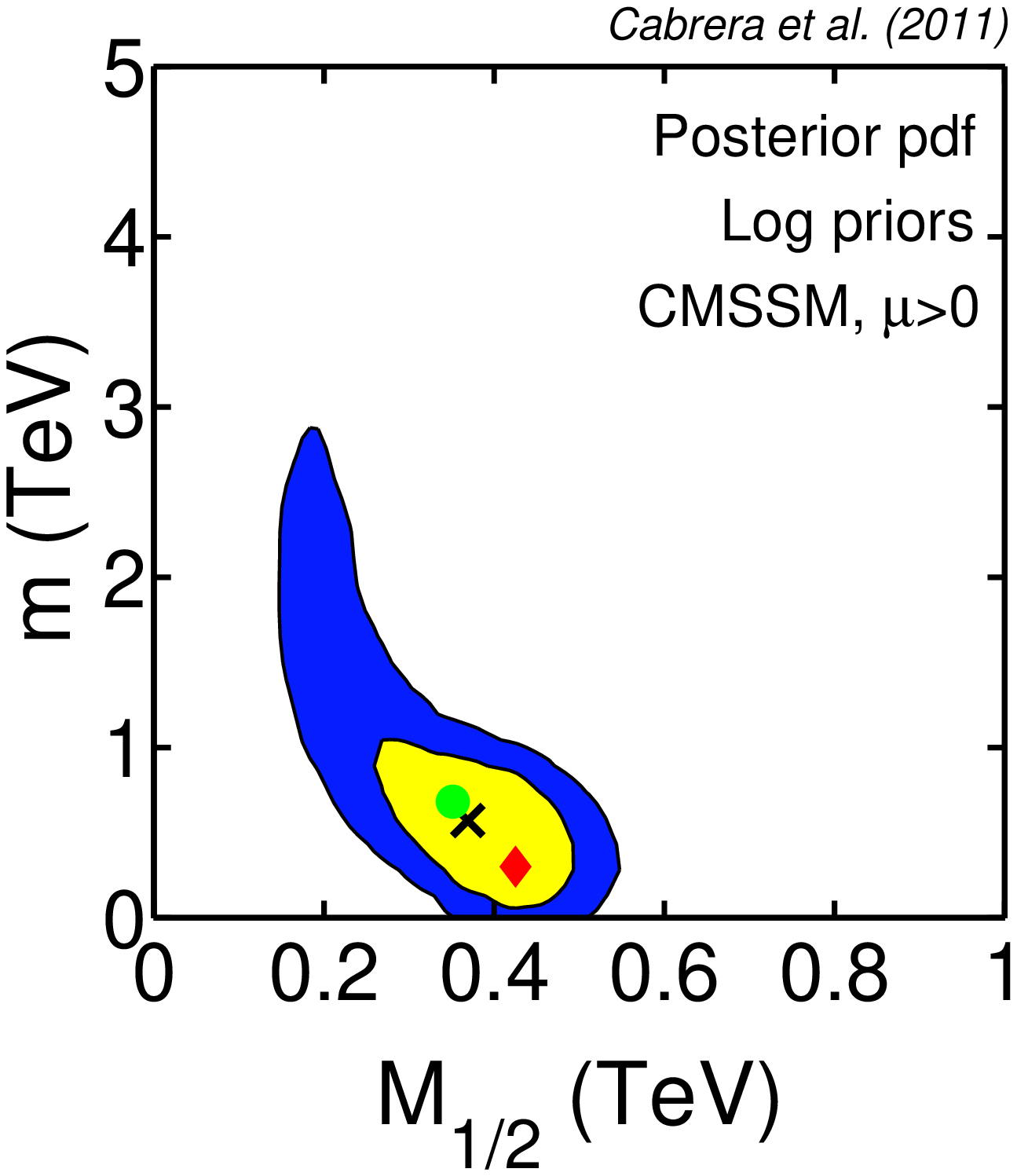} \\ 
\includegraphics[width=0.53\linewidth]{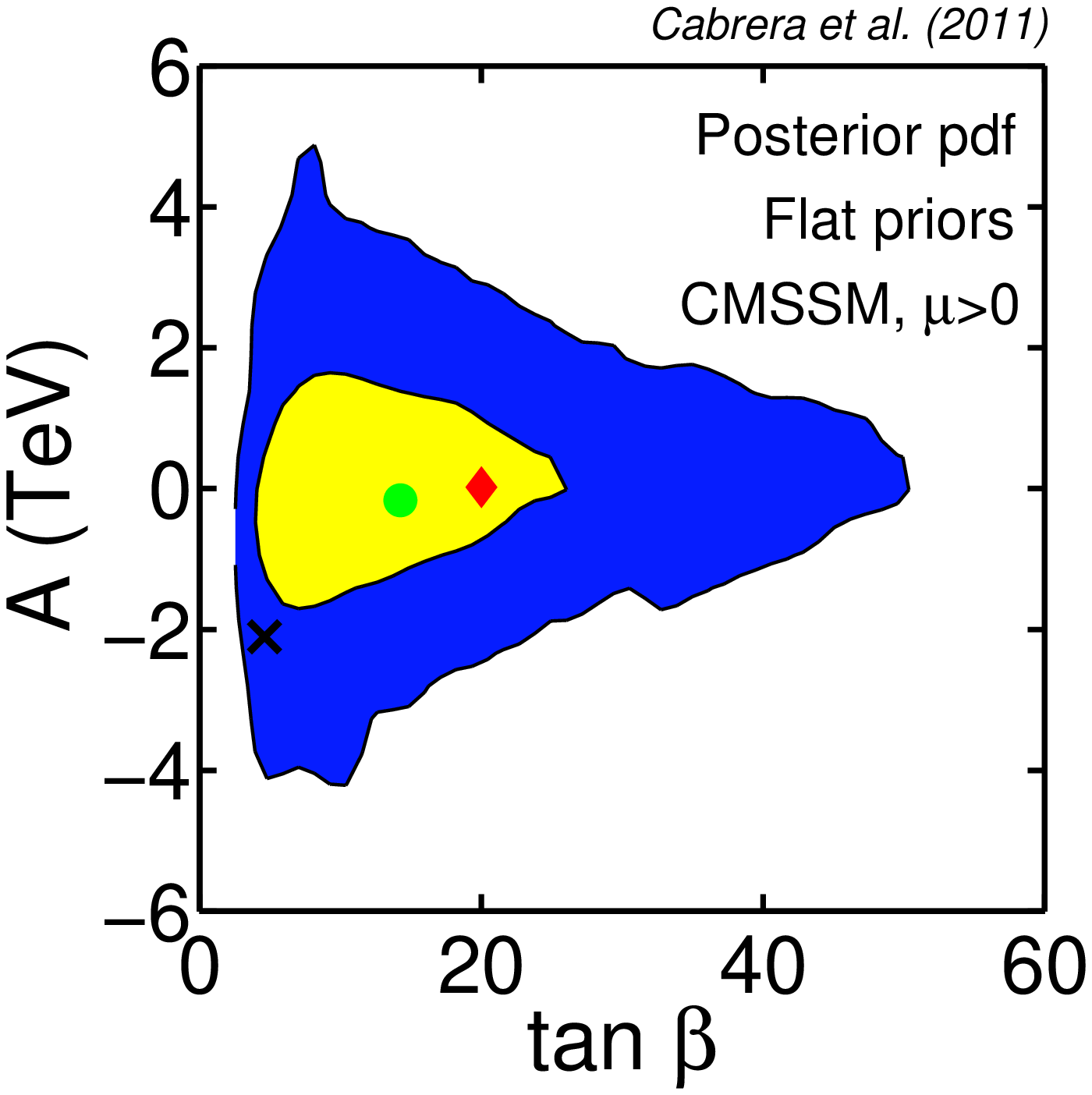} \hspace{-0.5cm}
\includegraphics[width=0.53\linewidth]{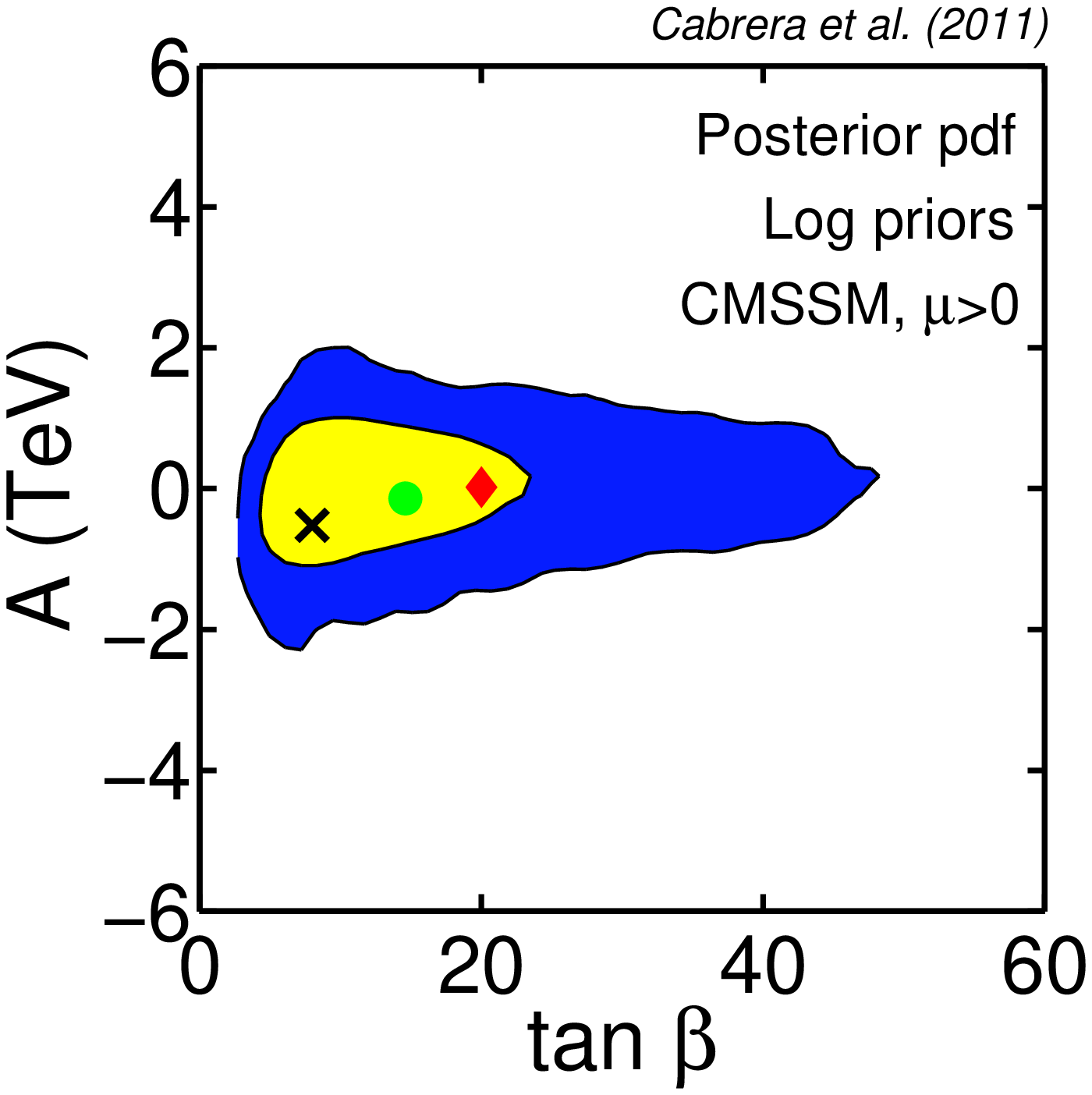}
\caption[text]{2D marginalized posterior probability density function for 
flat (left panels) and logarithmic (right panels) priors using the 
normalization test.  The inner and outer
contours enclose respective 68\% and 95\% joint regions. 
The small filled circle represents the mean value of the posterior pdf, 
the cross corresponds to the best-fit point found and the diamond to 
the SU9 model, used to produce the mock experimental data.} 
\label{fig:norm}
\end{figure}
\end{center}

Fig. \ref{fig:norm} (upper panels) shows the posterior pdf in the $M_{1/2}-m$
plane, after marginalizing the rest of the parameters: $A$, $\tan\beta$,
together with the previosly marginalized $\mu$ and the nuisance SM parameters,
$\{s\}$. As discussed below, the cross section for the kind of events
considered (multijets + missing transverse momentum) is actually fairly
insensitive to the values of $A$, $\tan\beta$, so the marginalization in these
parameters does not change appreciably the probability density in the
$M_{1/2}-m$ plane. The left (right) panel corresponds to log (flat) priors for
the soft terms. The shape of these plots can be easily understood. Since we
are fitting a unique quantity, namely the total number of events, and we have
two parameters, $\{M_{1/2}, m\}$, we can expect a degeneracy in the parameter
space, which is in fact the case. The elongated shape of the allowed region,
especially visible in the flat prior case, is in fact a widening --due to the
uncertainties-- of the line where the degeneracy is exact, which includes of
course the ``true model", i.e. SU9. This is marked with a red diamond in the
plots. Note that for small gaugino mass, the squark masses become irrelevant,
provided they are large enough, since in that case the dominant SUSY production are gluino pairs, whose masses do not depend on $m$. This is reflected in the vertical form of the region for small $M_{1/2}$.

\vspace{-1.5cm}
\noindent
Now, the shape of the line of degeneracy, somehow visible in the upper plots of Fig. \ref{fig:norm}, depends on the cuts used to select events. Note that, as the values of the soft terms get smaller, the cuts used to bound the energy of the first and second jets, and the missing transverse momentum (see previous subsection) become more and more inappropriate: many events with three or more jets plus missing transverse momentum do not pass the cuts. As a consequence the counted total number of events of this kind is dramatically cut out and can become equal to the experimental one. This enhances artificially the statistical weight of the low energy region. As a result the maximum value of the pdf, and its averaged central value (marked by a green dot), are shifted from the ``true model" (marked by a red diamond). 

There are ways to counteract these disagreeable effects. Playing with different cuts, the degeneracy gets partially broken and it is possible to discard larger regions of the parameter space. For instance, one can compare the total number of events using several choices for the lower bound on $M_{\rm eff}$. Somehow, this equivales to test the shape of the experimental and theoretical histograms, but not in the most efficient way. This is improved using the histogram-comparison technique explained in sections 2, 3, which we will apply shortly to this analysis.

Fig. \ref{fig:norm} (lower panels) shows the posterior in the $\tan\beta-A$ plane, after marginalizing the rest of the parameters. As mentioned above, the cross section of the type of events considered does not depend appreciably on $A$ and $\tan\beta$, and this is reflected in the plots. The preference for rather small values of both $A$ and $\tan\beta$ is essentially a consequence of the Jacobian factor (\ref{J_anal}) in the posterior (\ref{pdf2}). As commented in the previous subsection the Jacobian automatically penalizes regions of the parameter space where fine-tuning is needed to reproduce the electroweak scale. This disfavors large values for both $A$ and $\tan\beta$.\footnote{This is an statistical effect which is not visible in frequentist approaches, where the basic quantity is the likelihood and fine-tuning is not penalized, unless such penalization is artificially incorporated.} 
The remarkable insensitivity to $A$ and $\tan\beta$ is physically due to the fact that the CMSSM spectrum is not much dependent on the values of $A$ and $\tan\beta$, except for mixing effects in the mass matrices of stops (and sbottoms and staus for large $\tan\beta$), charginos and neutralinos. Even for these matrices the effect is normally quite small. Thus the production rates of squarks and gluinos  are quite independent of $A$ and $\tan\beta$. Once the supersymmetric particles are created, their decay rates are not very relevant for the cross section of the process considered (multijets + missing transverse momentum), and, in any case, they are quite independent of these parameters too. This insensitivity to $A$ and $\tan\beta$ could be partially cured by complementing the present analysis by a separate study of those events involving leptons \cite{Baer:1998sz}, but that discussion is outside the scope of this paper.

\begin{center}
\begin{figure}[t]
\hspace{-0.3cm}
\includegraphics[width=0.53\linewidth]{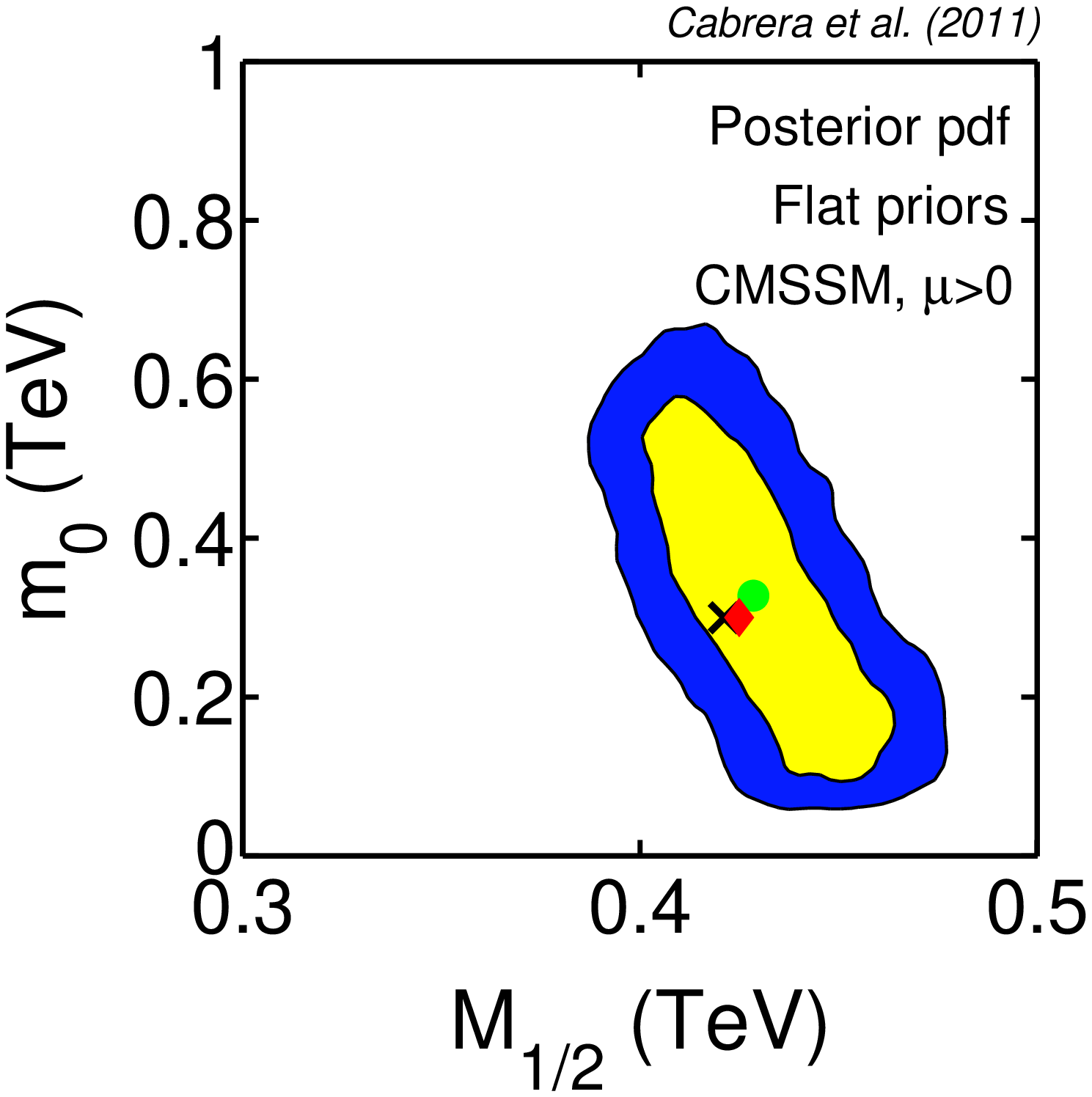} \hspace{-0.5cm}
\includegraphics[width=0.53\linewidth]{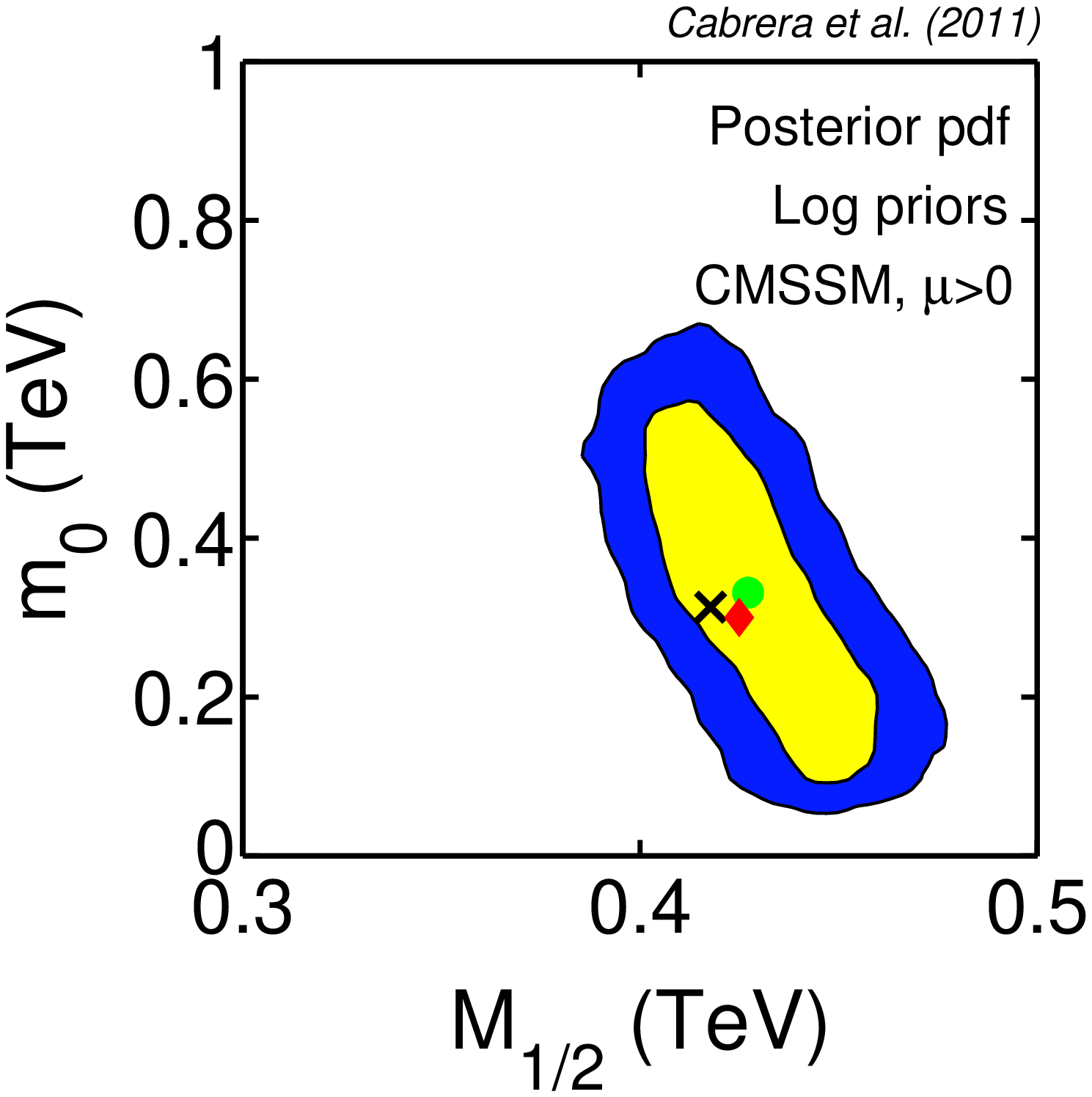} \\ 
\hspace{-0.3cm}
\includegraphics[width=0.53\linewidth]{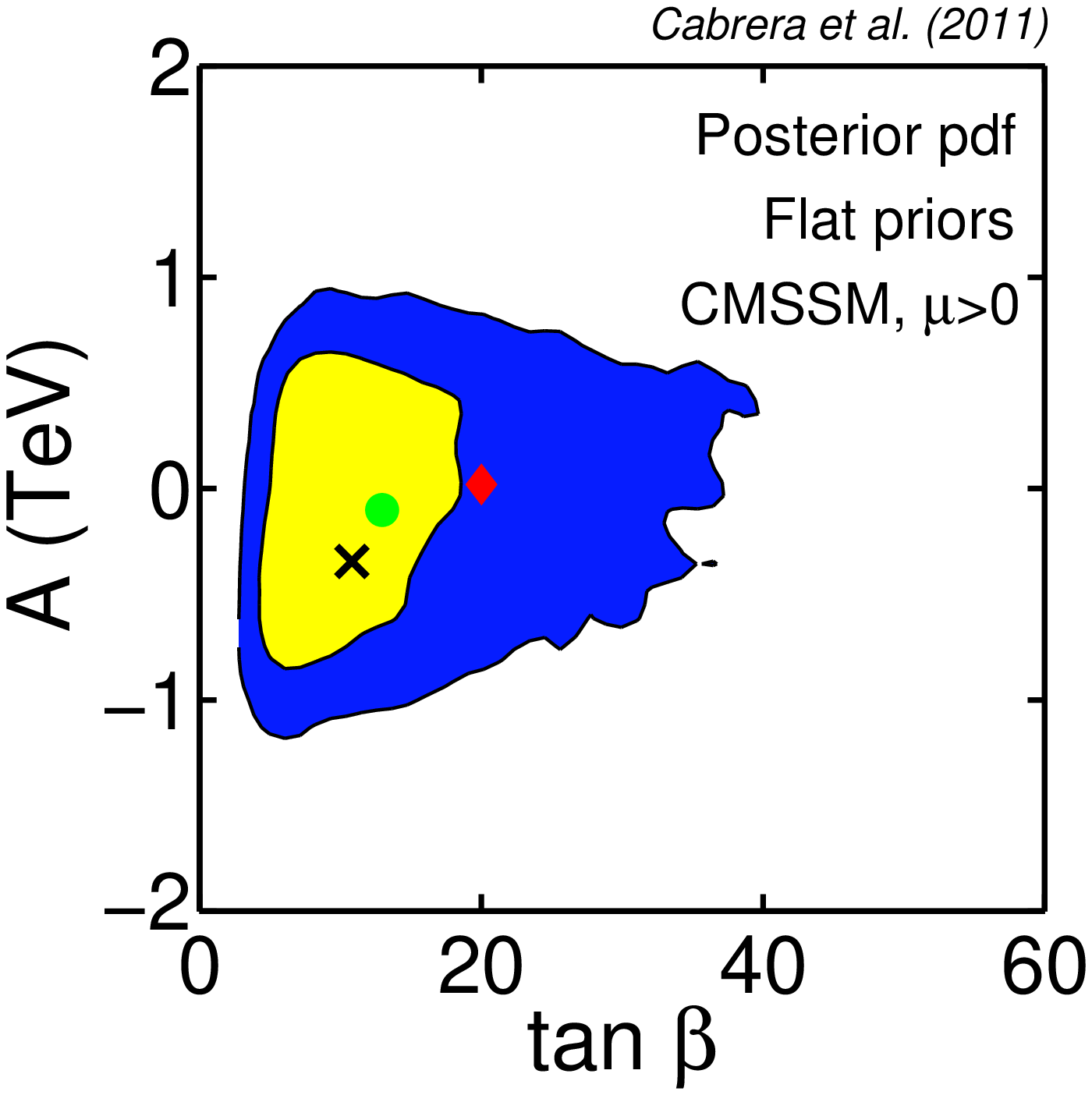} \hspace{-0.5cm}
\includegraphics[width=0.53\linewidth]{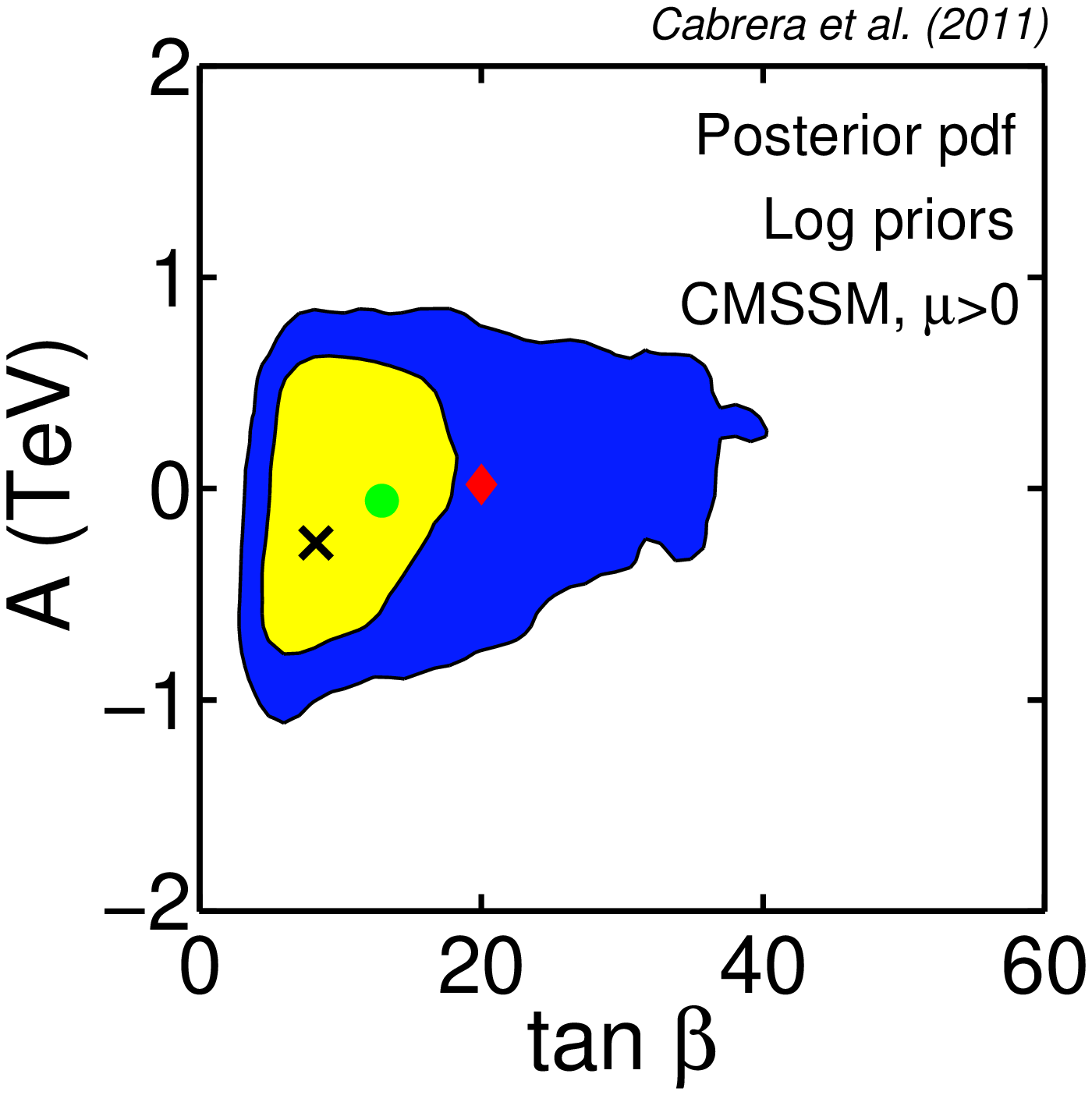}
\caption[text]{As Fig. 2 but using both normalization and shape. Note the
  different ranges of the two figures.} 
\label{fig:norm_shape}
\end{figure}
\end{center}

\vspace{-1.5cm}
\subsubsection*{Incorporation of the shape test}

Now we repeat the analysis, but computing the likelihood associated with the LHC data with the use of the whole expression (\ref{Pviui3}), which takes into account not only the total number of events, but also the comparison of the histogram shapes. Again, in order to incorporate the fact that the luminosity of the simulation changes from point to point in the parameter space we use the modified formula (\ref{Pviui3loggorro}):
\bea
{\rm LHC-likelihood}\ &\propto& \P(f=\frac{v}{Lu})
\nonumber\\
&\times&\
\prod_{i=1}^{K}\left( \frac{(u_i+v_i)!}{u_i! v_i!}\ \int dg_i \frac{1}{g_i}
\left(\frac{v}{u}g_i\right)^{v_i}\left(1+\frac{v}{u}g_i\right)^{-1-u_i-v_i}
\P(g_i)\right) \ .
\eea
We recall that $\P(g)$ carries all the systematic uncertainties affecting the shape of the histograms, and is given by the gaussian (\ref{Pg2}) with $\Delta_g=0.2$; as explained in the previous subsection. Note that, as could be expected, the correction due to the difference in luminosity does not affect the shape-part of the likelihood.

\begin{center}
\begin{figure}[t]
\includegraphics[width=0.5\linewidth]{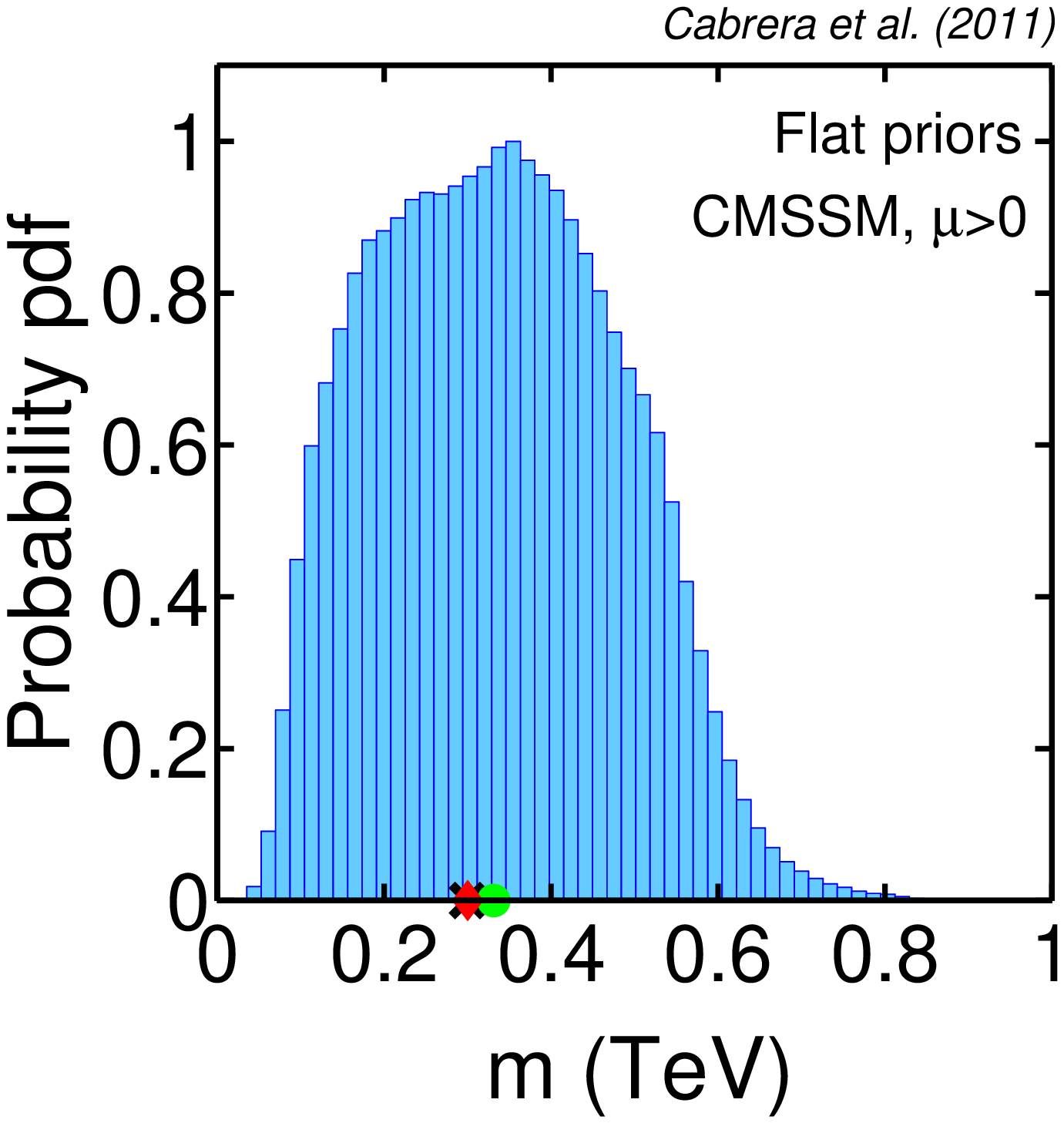} 
\includegraphics[width=0.5\linewidth]{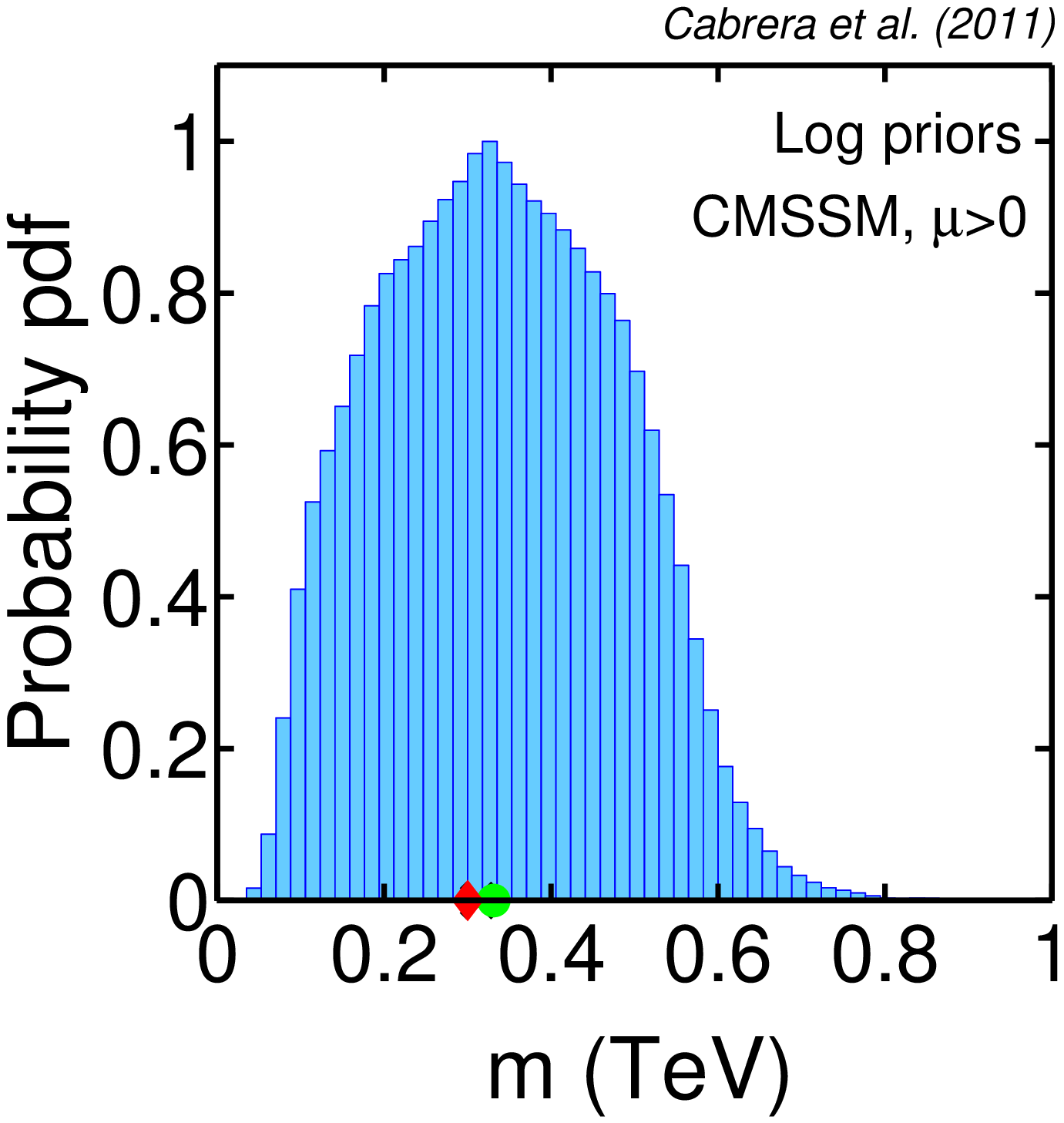} \\ 
\includegraphics[width=0.5\linewidth]{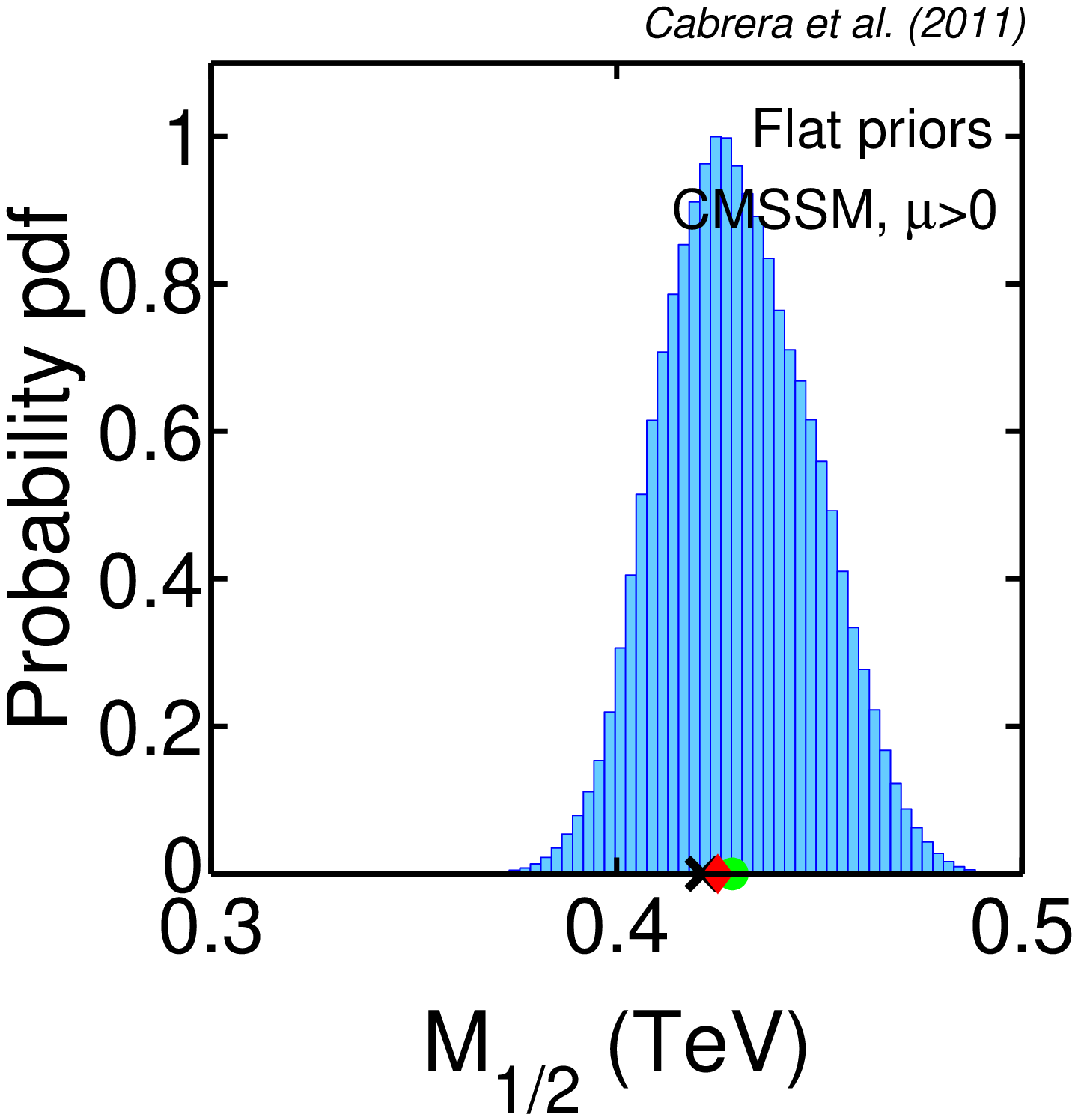} 
\includegraphics[width=0.5\linewidth]{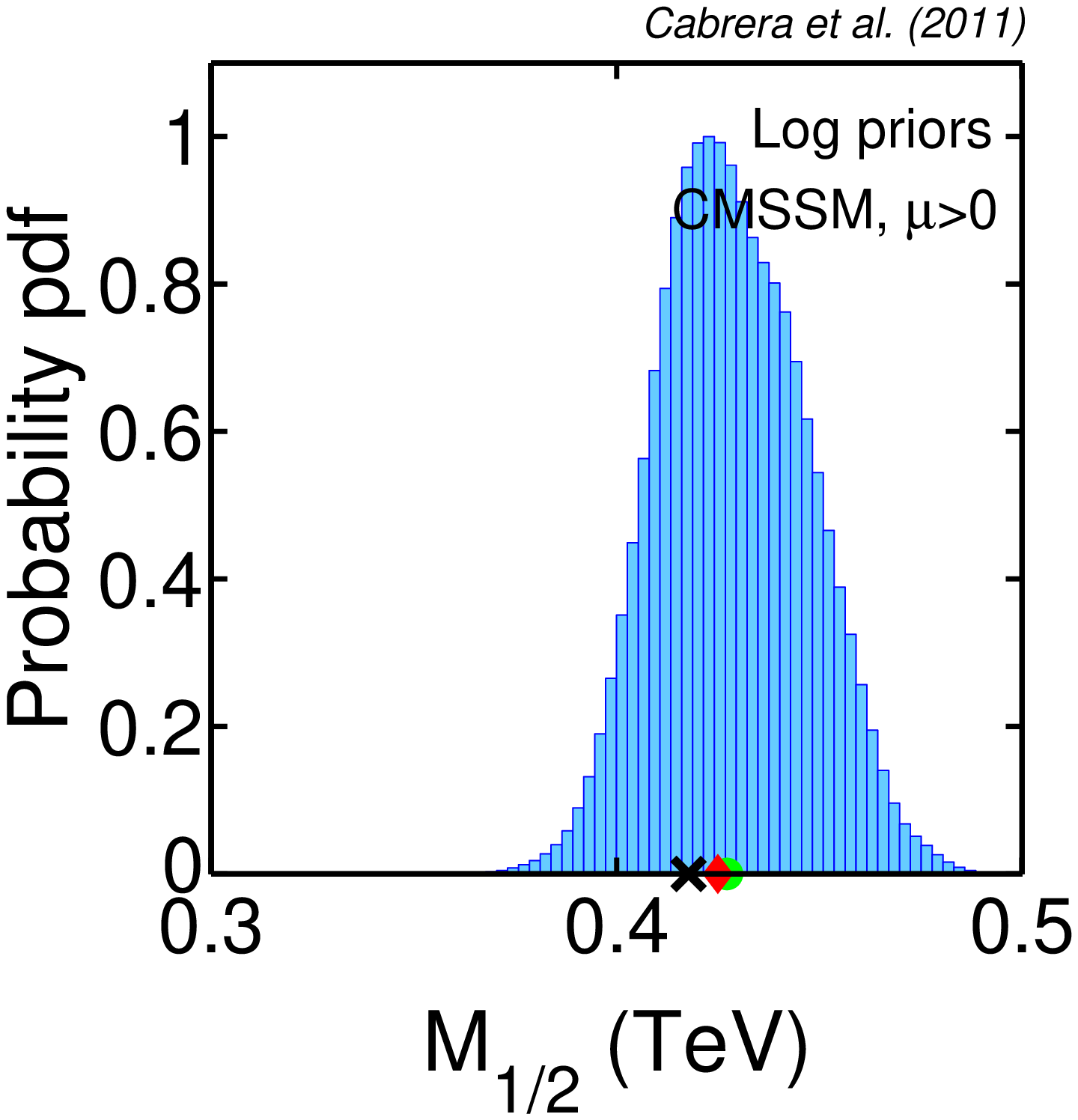}
\caption[text]{1D marginalized posterior probability density function of the 
$m$ and $M_{1/2}$ parameters (upper and lower panels respectively) for 
flat (left panels) and logarithmic (right panels) priors. 
The small filled circle represents the mean value of the posterior pdf, 
the cross corresponds to the best-fit point and the diamond to the SU9 model.} 
\label{fig:norm_shape_1D_1}
\end{figure}
\end{center}

\vspace{-1.5cm}
Fig. \ref{fig:norm_shape} is as Fig. \ref{fig:norm}, but after including the likelihood associated to the shape in the analysis. The upper panels show, for log and flat priors, the posterior pdf in the $M_{1/2}-m$ plane, after marginalizing the rest of the parameters. As expected, the test of the theory is now much more efficient and the previous degeneracies dissapear (note the different ranges of the two figures). This illustrates the potential of making use of all the information contained in the theoretical and experimental histograms when computing the likelihood of a model, provided the various sources of uncertainty are properly taken into account. 

The lower panels of Fig. \ref{fig:norm_shape} show the posterior in the $\tan\beta-A$ plane. Again, the cross section of the type of events considered does not depend appreciably on these parameters, which is reflected in the plots. Still, introducing the test for the shape slightly improves the sensitivity of the search to the values of $A$ and $\tan\beta$, but that sensitivity is anyway very small. 

\begin{center}
\begin{figure}[t]
\includegraphics[width=0.5\linewidth]{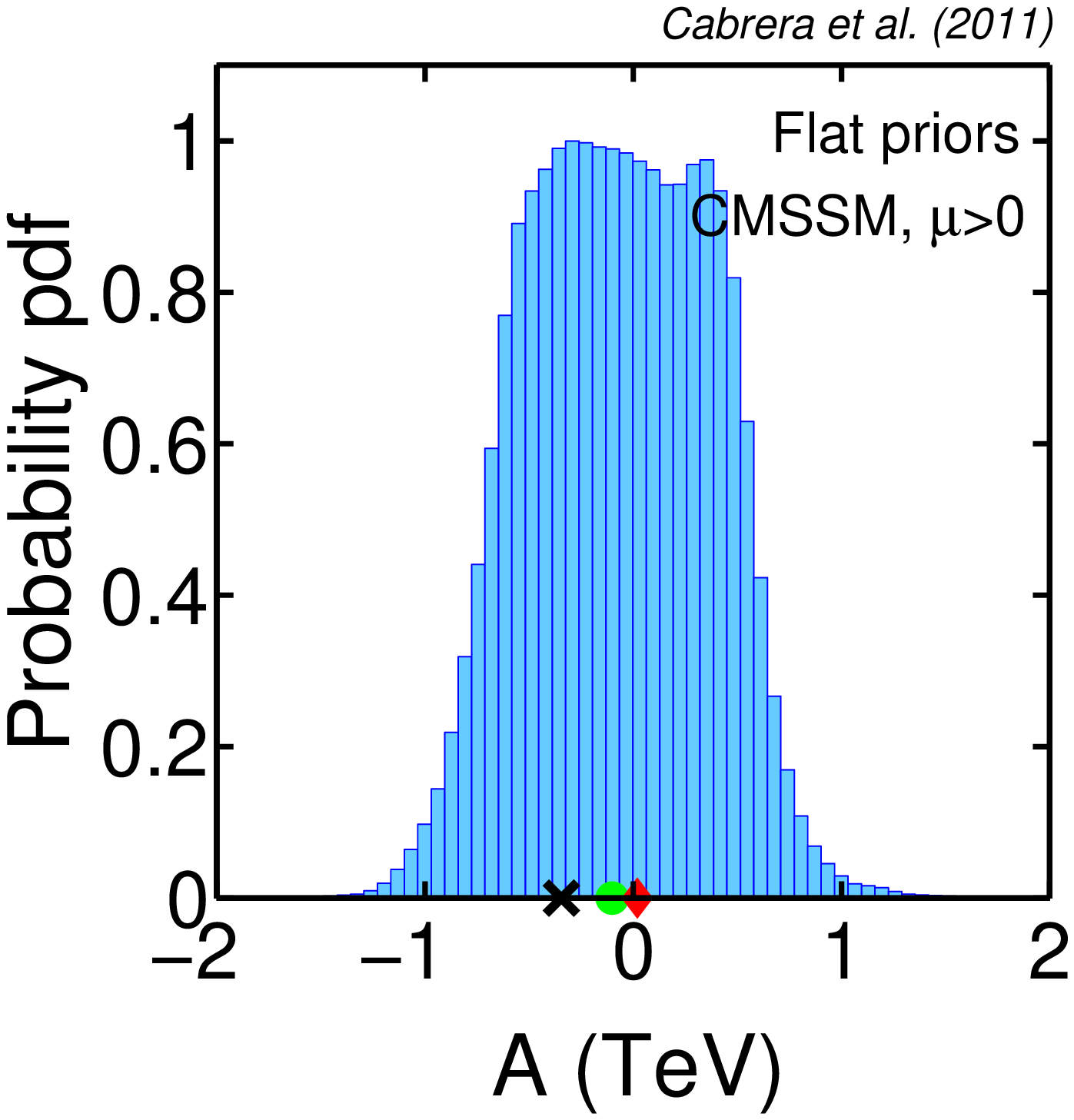} 
\includegraphics[width=0.5\linewidth]{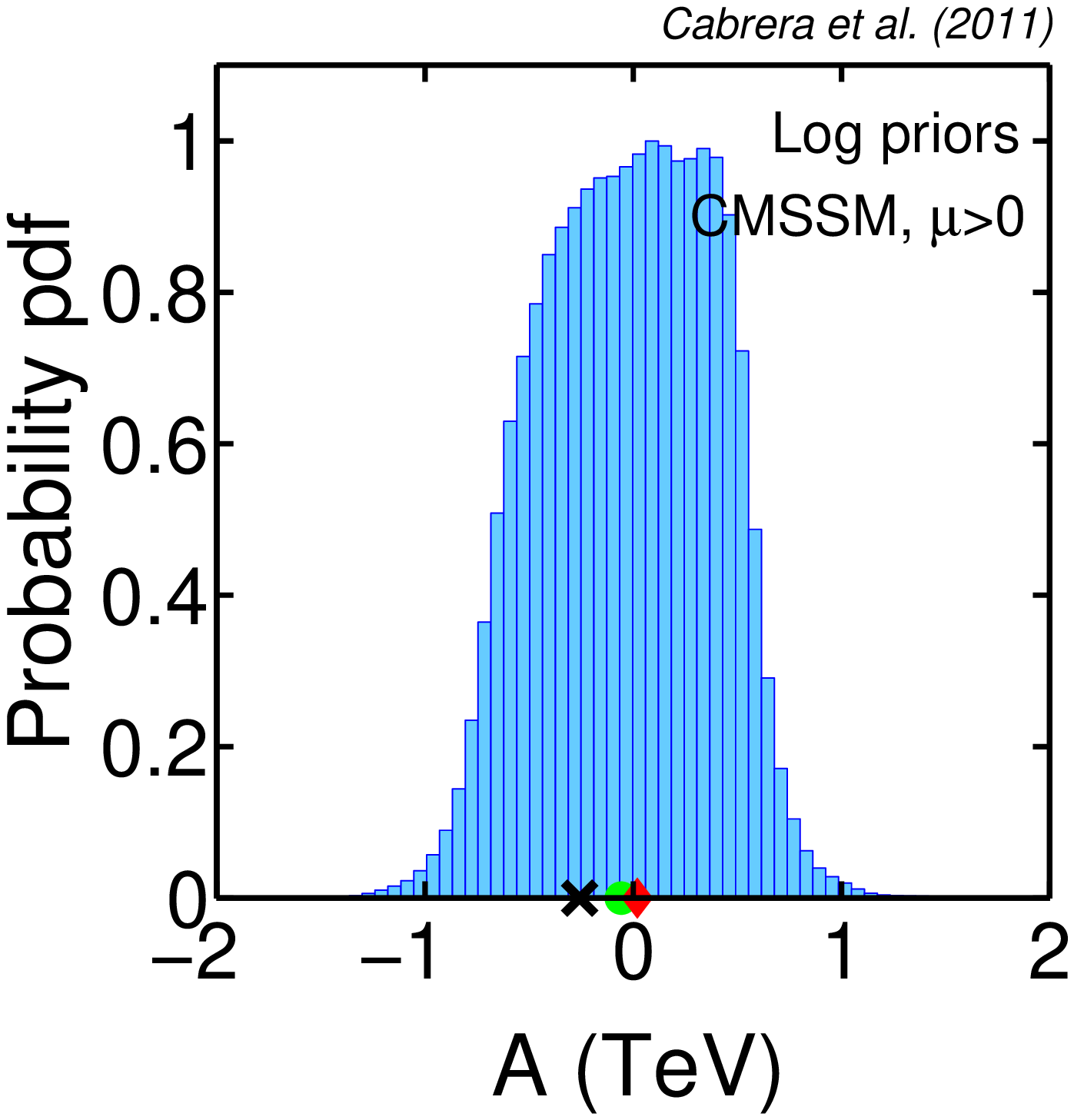} \\ 
\includegraphics[width=0.5\linewidth]{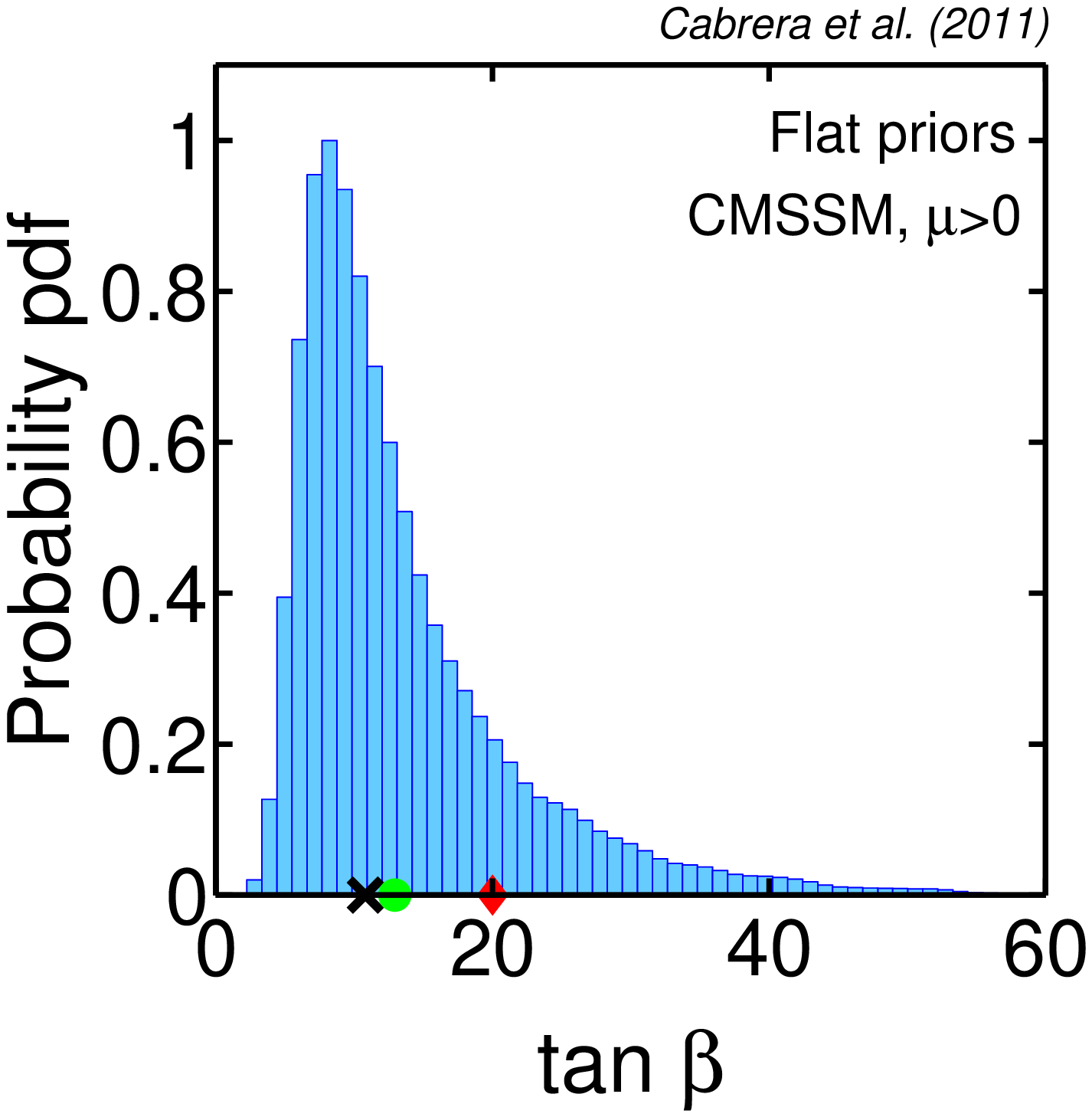} 
\includegraphics[width=0.5\linewidth]{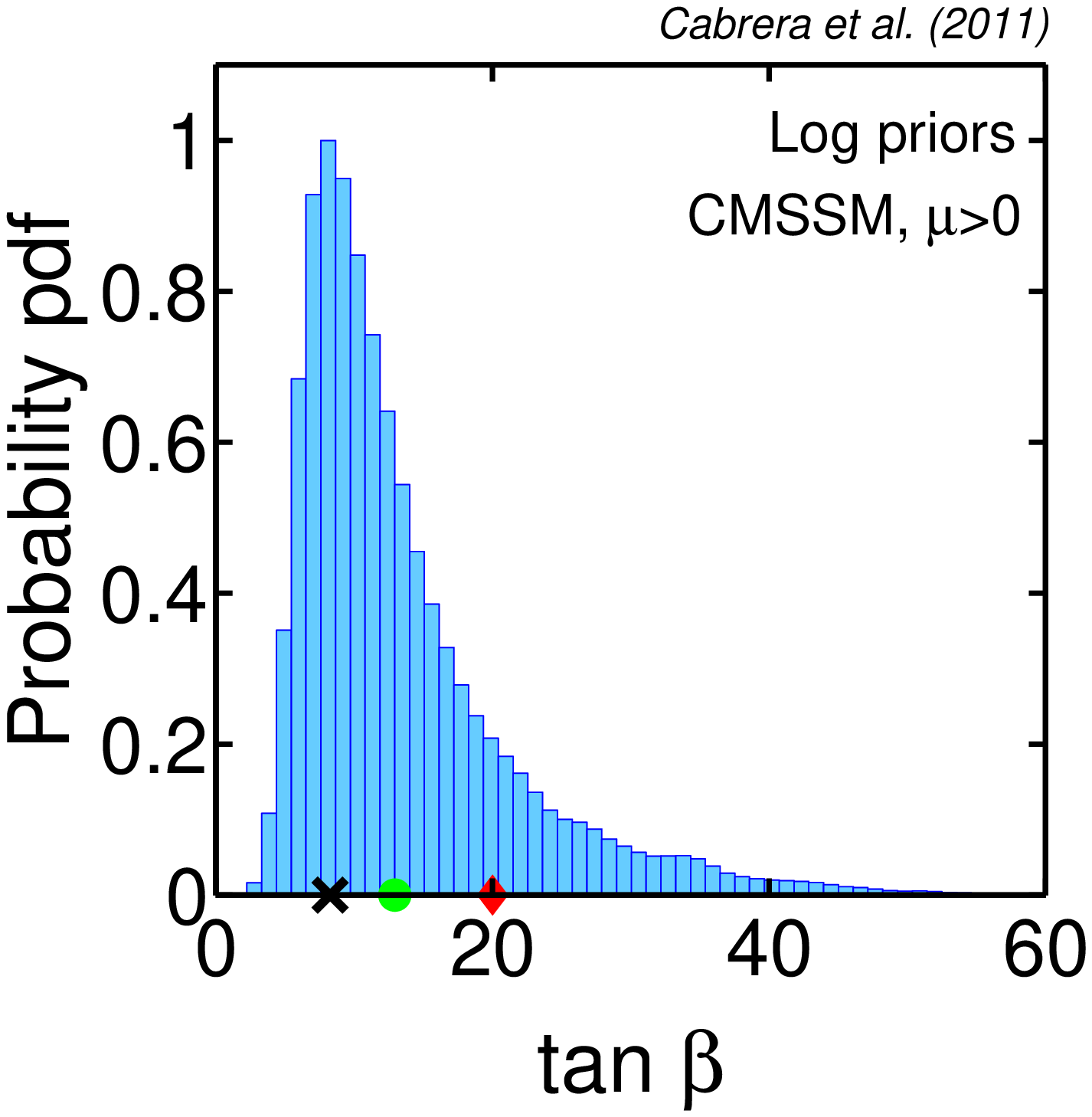}
\caption[text]{As Fig. 4 but for $A$ and $\tan\beta$ parameters.}
\label{fig:norm_shape_1D_2}
\end{figure}
\end{center}

Figs. \ref{fig:norm_shape_1D_1} and \ref{fig:norm_shape_1D_2} show, for logarithmic and flat priors, the unidimensional posteriors for $m$, $M_{1/2}$, $A$ and  $\tan\beta$, after marginalization of all the parameters, except the one plotted in each graph. The shape of these functions reflects the previous discussion. It is worth-noticing the great precision in the determination of the gaugino mass, which comes from the fact that, due to the renormalization group running, $M_{1/2}$ is the parameter that dominantly determines the low-energy spectrum of the CMSSM.

Finally, we note that the posteriors have in all cases a very slightly dependence on the type of prior used, reflecting the robustness of the approach.

\section{Conclusions}
\label{sec:summary}

Due to the complexity of the LHC experiment, much of the comparison between LHC data and theoretical predictions has to be made by confronting experimental histograms (in different variables) and theoretical histograms produced by simulations. In many cases the comparison is performed by comparing the total number of events after choosing a clever variable and applying convenient cuts. Other techniques make use of particular features of the histograms, like the presence of endpoints. But the procedure can be optimized by evaluating the actual likelihood associated to the complete histogram. The main goal of this paper has been precisely to present a rigorous and effective method to compare experimental and theoretical histograms, evaluating the total likelihood, and apply it to a physically relevant case. In doing this we have taken into account that, besides the statistical uncertainties inherent to the histograms, there are additional sources of systematic error. 

In the method presented, the complete likelihood is rigorously separated into two factors: the likelihood of the total number of events and the likelihood of the shape of the histogram. This in turn allows to treat the corresponding sources of systematic uncertainty in a separate way as well. This is very convenient when there are reasons to expect different systematic errors in the two pieces. The final formula for the total likelihood is given in eq.(\ref{Pviui3}). 

The procedure can be easily incorporated to both frequentist and Bayesian analyses, since both are based on the likelihood of the theoretical models. In the two approaches, incorporating the total likelihood optimizes the chances of picking up a signal of new physics and, once the signal is found, identifying which new physics is behind. E.g. if the new physics is supersymmetry, it allows to find in an optimal way the parameters of the supersymmetric model.

We have illustrated the latter point by showing how a search in the CMSSM parameter space, using Bayesian techniques, can effectively find the correct values of the CMSSM parameters by comparing histograms of events with multijets + missing transverse momentum displayed in the effective-mass variable. The procedure is in fact very efficient to identify the true supersymmetric model, in the case supersymmetry is really there and accessible to the LHC. But, of course, the technique can be applied to any scenario of new physics.

\section{Appendix A: Histogram comparison when experiment and the simulation have different luminosities}

Some expressions of sects. 2 and 3 have to be modified when the effective luminosity of the simulation is not the same as the experimental one. In practice, the former can change from point to point when scanning the parameter space since typically one simulates a fixed number of supersymmetric events (say $10^4$ events), but obviously the cross section changes throughout the parameter space. Of course one could adjust at every point the luminosity so that it coincides with the experiment, but normally this is costly in running time, and it is unnecessary, since the comparison can still be made as described next.

Let us call $L^{th}$, $L^{exp}$ the luminosities of the theoretical simulation and the experiment, respectively, and suppose for a moment there are no systematic errors. 
Then the means that, under the null-hypothesis, are responsible for the
experimental data, $v_i$, are {\em not} the ones of the simulation, say $\hat{\mu}_i$, but
\bea
\label{muthgorro}
{\mu}_i=\frac{L^{exp}}{L^{th}}\ \hat{\mu}_i\ \equiv\ L\ \hat{\mu}_i \ .
\eea
Hence eq.(\ref{nosys3}) becomes
\bea
\label{nosys3gorro}
P(v_i|u_i)=\int \prod_{i=1}^{K}\ d\hat{\mu}_i \frac{(L \hat{\mu}_i)^{v_i}}{v_i!}e^{-L \hat{\mu}_i}\ \frac{\hat{\mu}_i^{u_i}}{u_i!}e^{-\hat{\mu}_i}=\prod_{i=1}^{K} \frac{(u_i+v_i)!}{u_i!\ v_i!}\ \ L^v\ (1+L)^{-1-u_i-v_i} \ .
\eea
Once systematic uncertainty is taken into account, see section \ref{sec:errors}, everything is actually easy to handle since the luminosity factor $L$ plays the role of a systematic and universal factor affecting the means of the simulation. More precisely, the equation (\ref{mumuth}), that relates the true means to be compared with the experiment, $\mu_i$, with those of the simulation, $\muth_i$, becomes
\bea
\label{mumuthgorro}
\mu_i \ =\ L\ f\ g_i\ \muth_i \ .    
\eea
Therefore the subsequent equations remain the same with the simple change $f\rightarrow Lf$.
In particular, the likelihood given by eq.(\ref{Pviui3}) becomes now
\bea
\label{Pviui3loggorro}
\P(v_i|u_i)\propto \P(f=\frac{v}{Lu})
\ \prod_{i=1}^{K}\left( \frac{(u_i+v_i)!}{u_i! v_i!}\ \int dg_i \frac{1}{g_i}
\left(\frac{v}{u}g_i\right)^{v_i}\left(1+\frac{v}{u}g_i\right)^{-1-u_i-v_i}
\P(g_i)\right) \ .
\eea
This is the formula we have used in our scan of the CMSSM parameter space.

\section*{Acknowledgements}

We thank F. Feroz and R. Trotta for interesting discussions and suggestions. \\
This work has been partially supported by the MICINN, Spain, under contract FPA 2007--60252, the Comunidad de Madrid through Proyecto HEPHACOS S-0505/ESP--0346, and by the European Union through the UniverseNet (MRTN--CT--2006--035863).
M.~E. Cabrera acknowledges the 
financial support of the CSIC through a predoctoral research grant (JAEPre 07 00020).
The work of R. Ruiz de Austri has been supported in part by the project 
PARSIFAL (FPA2007-60323) of the Ministerio de Educaci\'{o}n y 
Ciencia of Spain and by the Spanish MICINN's Consolider-Ingenio 2010 
Programme under the grant MULTIDARK CSD2009-00064, CPAN CSD2007-00042 and PAU CSD2007-00060.
V.A.M. acknowledges support by the Spanish Ministry of Science and Innovation (MICINN) under the project FPA2009-13234-C04-01, by the Ram\'on y Cajal contract RYC-2007-00631 of MICINN and CSIC, by the Spanish Agency of International Cooperation for Development under the PCI project A/030322/10 and by the grant UV-INV-EPDI11-42955 of the University of Valencia.
The use of the Hydra cluster of the IFT-UAM/CSIC is also acknowledged.


\end{document}